\documentclass[
  twocolumn,
]{aastex631}

\newcommand{\handleAckLineNumbers}{}  

\usepackage{amsmath}

\newcommand{\customhalfwidth}{
  \ifdim 2\linewidth>\textwidth   
    0.5\textwidth
  \else   
    \linewidth
  \fi
}

    \newcommand{\pdfrac}[2]{\frac{\partial #1}{\partial #2}}  
    \newcommand{\advect}[2]{\frac{d_{#2} #1}{dt}}             
    \newcommand{\pdtime}[1]{\frac{\partial #1}{\partial t}}   
    \newcommand{\inlineAdvect}[2]{d_{#2} #1 / dt}
    \newcommand{\inlinePdtime}[1]{\partial #1 / \partial t}
    \newcommand{\grad}[1]{\nabla #1}                          
    \newcommand{\divg}[1]{\nabla \cdot #1}                    
    \newcommand{\curl}[1]{\nabla \times #1}                   
    
    \newcommand{\eqdef}{=}    

    \newcommand{\enc}[3]{\left#1#2\right#3}
    
    \newcommand{\parens}[1]{\enc{(}{#1}{)}}
    \newcommand{\brackets}[1]{\enc{[}{#1}{]}}

    \newcommand{\W}{\omega}    
    \newcommand{\K}{\vec{k}}   
    \newcommand{\oZsym}{(0)}     
    \newcommand{\oZ}{^{\oZsym}}  

    \newcommand{\A}{s}  
    \newcommand{\B}{j}   
    \newcommand{\C}{\alpha}       
    \newcommand{\D}{\beta}       
    \newcommand{\Wterm}{\mathcal{W}}
    \newcommand{\Ldebye}[1]{\lambda_{D, #1}}
    \newcommand{\Fterm}{\mathcal{G}}

  \newcommand{\TFBI}{Thermal Farley-Buneman instability}  
  \newcommand{\TFBIabbrv}{TFBI}
  
  \newcommand{\TFBIabbrvdefp}{Thermal Farley-Buneman instability (TFBI)}
  \newcommand{\IRIS}{\textit{IRIS}}
  \newcommand{\ALMA}{\textit{ALMA}}

  \newcommand{\code}[1]{\texttt{\detokenize{#1}}}

  \newcommand{\BUCSP}{Boston University Center for Space Physics, 725 Commonwealth Ave, Boston, MA 02215, USA}

\begin{document}

\title{Multi-fluid Simulation of Solar Chromospheric Turbulence and Heating Due to the Thermal Farley-Buneman Instability}

  \author[0000-0002-1127-7350]{Samuel Evans}
    \affiliation{\BUCSP}
  \author[0000-0002-8581-6177]{Meers Oppenheim}
    \affiliation{\BUCSP}
  \author[0000-0002-0333-5717]{Juan Mart\'inez-Sykora}
    \affiliation{Lockheed Martin Solar \& Astrophysics Laboratory, 3251 Hanover St, Palo Alto, CA 94304, USA}
    \affiliation{Bay Area Environmental Research Institute, NASA Research Park, Moffett Field, CA 94035, USA}
    \affiliation{Rosseland Centre for Solar Physics, University of Oslo, P.O. Box 1029 Blindern, N-0315 Oslo, Norway}
    \affiliation{Institute of Theoretical Astrophysics, University of Oslo, P.O. Box 1029 Blindern, N-0315 Oslo, Norway}
  \author[0000-0002-3807-5820]{Yakov Dimant}
    \affiliation{\BUCSP}
  \author{Richard Xiao}
    \affiliation{\BUCSP}

\begin{abstract}
  Models fail to reproduce observations of the coldest parts of the Sun's atmosphere, where interactions between multiple ionized and neutral species prevent an accurate MHD representation. This paper argues that a meter-scale electrostatic plasma instability develops in these regions and causes heating. We refer to this instability as the Thermal Farley-Buneman instability, or TFBI. Using parameters from a 2.5D radiative MHD Bifrost simulation, we show that the TFBI develops in many of the colder regions in the chromosphere. This paper also presents the first multi-fluid simulation of the TFBI and validates this new result by demonstrating close agreement with theory during the linear regime. The simulation eventually develops turbulence, and we characterize the resulting wave-driven heating, plasma transport, and random motions. These results all contend that effects of the TFBI contribute to the discrepancies between solar observations and radiative MHD models.
\end{abstract}

\section{Introduction} \label{sec:intro}

  The chromosphere is the complex interface region between the photosphere and the million-degree corona. For solar modeling it is crucial to understand the chromosphere, since all energy transfer from the surface of the Sun to the corona must pass through this intermediary region. The chromosphere presents a modeling challenge as it spans many parameter regimes, microphysics may play an important role, and the assumptions of MHD break down. Over the last few decades, large improvements have been made with radiative (M)HD models, which capture a large variety of chromospheric dynamics such as magneto-acoustic shocks \citep[see e.g.][]{Carlsson1992, Carlsson1995, Carlsson2002, Wedemeyer2004, Carlsson2007}, spicules \citep{Hansteen2007, MartinezSykora2017}, and flux emergence \citep{Cheung2014}. Some models have been further improved to include the effects of ion-neutral interactions \citep[][and references therein]{Leake2014, MartinezSykora2015, Ballester2018} 
  and non-equilibrium ionization \citep{Leenaarts2007, Golding2014, Przybylski2022}.

  However, comparisons between chromospheric observables and synthesis from those models reveal large discrepancies in some areas. The observed profiles, such as Mg~II from \IRIS\ \citep{DePontieu2014}, are typically wider than the corresponding synthesized profiles \citep{Carlsson2019}. Additionally, comparisons between \IRIS\ and \ALMA\ observations and recent single-fluid radiative MHD models, which include ion-neutral interaction and non-equilibrium effects, indicate that spicules are up to a few thousand degrees colder in the models \citep{Chintzoglou2021}. These issues might be alleviated by including the heating and turbulent effects of small-scale plasma instabilities, which are missing from such models.

  Fontenla et al. \citep{Fontenla2005, Fontenla2008} originally suggested that the Farley-Buneman (FB) instability can lead to heating in the chromosphere. 
  They argued that convective motions of neutral gas drag the mostly collisionally-demagnetized ions across the solar magnetic field while the electron motion remains primarily tied to the magnetic field lines.
  This causes the development of currents and electrostatic waves which lead to instability, as described in \citet{Dimant1995}. The work by \citet{Fontenla2005, Fontenla2008} treats the FB instability appropriately for the ionosphere, where it was originally discovered, but neglects crucial terms that become relevant in the chromosphere. \citet{Madsen2014} includes some such terms by treating the instability with a multi-fluid model, yet neglects proton magnetization; \citet{Fletcher2018} shows that ion magnetization effects reduce the prevelance of the instability in the chromosphere.

  By simulating the instability using a kinetic Particle-in-Cell code, \citet{Oppenheim2020} discovered that temperature perturbations significantly affect instability properties under chromospheric parameters. They improved the theory to include these thermal effects, predicting the new instability which we refer to as the \TFBIabbrvdefp. \citet{Dimant2022.TFBI} further studies the linear theory of this instability in different limiting cases, and determines that a multi-fluid model may be sufficient to reproduce the \TFBIabbrv\ for chromospheric parameters.

  The \TFBI\ is not limited to only appearing in the solar chromosphere, though. It could appear in other stellar atmospheres as well, and likely appears in various planetary ionospheres, including Earth's ionosphere. It may affect the dynamics in molecular clouds, and heat transfer in accretion disks. Any partially ionized plasma having sufficiently strong flows across magnetic field lines, along with the appropriate fluid densities and temperatures, may produce the \TFBIabbrv.

  To study the \TFBIabbrv\ in the Sun's chromosphere, we utilize the new multi-fluid code, Ebysus \citep{MartinezSykora2020.Ebysus}. Ebysus treats each ionized level of each atomic species as a separate fluid, with the ability to handle any number of fluids in the same simulation. Using Ebysus, we simulate the multi-fluid \TFBI\ in a fluid-model code for the first time. The ability to produce this instability in a fluid-model code enables studies at larger scales and across more chromospheric parameter regimes than what is possible with kinetic codes alone. 

  The remainder of this paper is structured as follows. Section~\ref{sec:theory and model} discusses the instability theory and simulation setup. Section~\ref{sec:theory results} describes our prediction about where the multi-fluid \TFBIabbrv\ occurs in the Sun's chromosphere, based on a single-fluid radiative 2.5D simulation of the solar atmosphere. Section~\ref{sec:simulation results} shows our multi-fluid simulation output and confirms the growth rate and wave properties agree with the \TFBIabbrv\ theory. Section~\ref{sec:heating} discusses the non-linear stage of the simulation and the resulting heating and random motions. This paper concludes with a summary of the results in Section~\ref{sec:conclusion}. 

\section{Theory and Simulation Structure} \label{sec:theory and model}
  Both the instability theory \citep[][see also Appendix~\ref{sec:disprel}]{Dimant2022.TFBI} and the Ebysus \citep{MartinezSykora2020.Ebysus} simulations in this work use multi-fluid models to study the chromosphere. In these models, the continuity, momentum, and energy equations govern the number density, $n_\A$; velocity, $\vec{u}_\A$; and temperature (in energy units), $T_\A$, for each fluid ($\A$):
  \begin{subequations} \label{eqs:fluid}
    \begin{widetext}
      \begin{equation} \label{eq:continuity}
        \pdtime{n_\A} + \divg{n_\A \vec{u}_\A} = 0
      \end{equation}
      \begin{equation} \label{eq:momentum}
        n_\A \advect{\vec{u}_\A}{\A}
        =
        - \frac{\grad{\parens{n_\A T_\A}}}{m_\A}   
        + n_\A \frac{q_\A}{m_\A} \parens{ \vec{E} + \vec{u}_\A \times \vec{B} }
        + \sum_\B n_\A \nu_{\A\B} \parens{ \vec{u}_\B - \vec{u}_\A }
      \end{equation}
      \begin{equation} \label{eq:heating}
        \advect{T_\A}{\A}
        =
        - \frac{2}{3} T_\A \divg{\vec{u}_\A}
        + \sum_\B
          \frac{2 m_\A}{m_\A + m_\B} \nu_{\A\B} \brackets{
              \frac{m_\B}{3} \left( \vec{u}_\B - \vec{u}_\A \right)^2
              + \left( T_\B - T_\A \right)
          }
      \end{equation}
    \end{widetext}

    \noindent where $\inlineAdvect{f}{\A} \eqdef \inlinePdtime{f} + \vec{u}_\A \cdot \grad{f}$, and
    sums are taken over all fluids including electrons. The atomic mass and charge of fluid species $\A$ are 
    $m_\A$ and $q_\A$, while $\vec{E}$ and $\vec{B}$ are the electric and magnetic fields, respectively. The collision frequency for momentum transfer to fluid $\A$ from fluid $\B$ is $\nu_{\A\B}$, for $\A \neq \B$, and the models assume elastic collisions. 
    Note that these models treat each species as an ideal gas, 
    and neglect the effects of ionization and recombination, thermal conduction, and gravity.

    The models also assume quasineutrality:
    \begin{equation} \label{eq:QN}
      \sum_\A n_\A q_\A = 0
    \end{equation}
    and use it, instead of the electron continuity equation, to solve for $n_e$.
  \end{subequations}

  The theory closes the system via the electrostatic assumption, $\inlinePdtime{\vec{B}}=0$.
  The mean electric field can be determined from the electron momentum equation, while perturbations of $\vec{E}$ are handled by the linear theory.

  Meanwhile, the Ebysus code allows $\vec{B}$ to vary. Ebysus includes equations \eqref{eqs:fluid}, as well as Faraday's and Ampere's laws without displacement current:
    \begin{subequations} \label{eqs:maxwell}
      \begin{equation} \label{eq:curlE}
        \pdtime{\vec{B}} = - \curl{\vec{E}}
      \end{equation}
      \begin{equation} \label{eq:curlB}
        \mu_0 \vec{J} = \curl{\vec{B}}
      \end{equation}
    \end{subequations}
  \noindent where $\mu_0 \approx 4 \pi \times 10^{-7} N/A^2$ is the vacuum permeability constant, and $\vec{J}\eqdef \sum_\A n_\A q_\A \vec{u}_\A$ is the current.

  Ebysus determines $\vec{B}$ by updating it every timestep using Faraday's law \eqref{eq:curlE}. To calculate the electric field, Ebysus solves the electron momentum equation \eqref{eq:momentum} for $\vec{E}$, assuming negligble electron inertia: $(m_e / q_e) \inlineAdvect{\vec{u}_e}{e} \rightarrow 0$. Finally, to determine the electron velocity, Ebysus solves for $\vec{u}_e$ using Ampere's law \eqref{eq:curlB} and the definition of current. This fully closes the system of equations in the Ebysus model.

  In the next sections we discuss the instability theory, the initial conditions of the multi-fluid Ebysus simulation presented in this work, and the numerical methods utilized by Ebysus.

  \subsection{Linear Theory of the Thermal Farley-Buneman Instability} \label{sec:theory}

    Linear instability theory makes predictions about small perturbations in a static background. This theory considers plane waves with real wavevector $\K$ and complex frequency $\W$, where all perturbations are proportional to $\exp\brackets{i\parens{\K\cdot\vec{x} - \W t}}$. Solutions with $\text{Im}(\W) > 0$ are unstable, with exponential growth rate $\text{Im}(\W)$. 

    This paper applies the linear theory described in \citet{Oppenheim2020, Dimant2022.TFBI}. In addition to the effects present in the Farley-Buneman instability, this work includes physical effects relevant in the chromosphere: thermal perturbations, arbitrary ion and electron magnetization, and generalizing for arbitrarily many ion fluids. 
    
    Beyond equations \eqref{eqs:fluid}, we make some additional assumptions to simplify the algebra. In particular, we assume the plasma is weakly ionized and contains only one neutral fluid, $n$, which does not respond to any perturbations, and we neglect Coulomb collisions while assuming all other collision frequencies are constant. The weak ionization assumption is reasonable in extended regions of the chromosphere, and in those regions the Coulomb collisions are orders of magnitude smaller than collisions with neutrals \citep[][and references therein]{Wargnier2022}. We also assume the perturbation is electrostatic, i.e., any magnetic field response to the perturbation is negligible. Additionally, since two dimensions perpendicular to $\vec{B}$ are sufficient to reproduce the \TFBIabbrv, we consider only such solutions here, enforcing $\K \cdot \vec{B}=0$. Finally, we assume that the unperturbed values are constant in space and time. This assumption of vanishing gradients may need revisiting in the future, as gradients might provide important contributions to the instability in some parameter regimes. The equations and assumptions above lead to the dispersion relation for this model, summarized in Appendix~\ref{sec:disprel}.
    
    Figure~\ref{fig:growthrate vs k theory} shows the predicted linear instability growth rate for the set of parameters in Table~\ref{tab:simulation params}, representing a cold region from a simulated chromosphere (see Section~\ref{sec:theory results}). At each wavevector $\K$, the growth rate is the largest imaginary part of all the solutions $\W$ to the dispersion relation (Eqs.~\ref{eqs:disprel}). This prediction is calculated numerically by converting the dispersion relation to a polynomial in $\W$ and applying a polynomial root-finding algorithm; see Appendix~\ref{sec:disprel} for more details.
    The maximum growth rate of roughly $\text{Im}(\W_{\text{peak}}) = 2440$~$s^{-1}$ occurs at $\K_{\text{peak}} = (6.83 \hat{x} - 0.94 \hat{y})$~$m^{-1}$. This corresponds to plane waves with a wavelength of 0.91~$m$, at an angle of roughly $8\degr$ below the $+x$ axis. This angle is $65\degr$ counterclockwise from the $\vec{E}$ direction, and the magnitude of the electric field is $|\vec{E}|=8.9$~V/m.

    \begin{figure}  
      \includegraphics[width=\customhalfwidth]{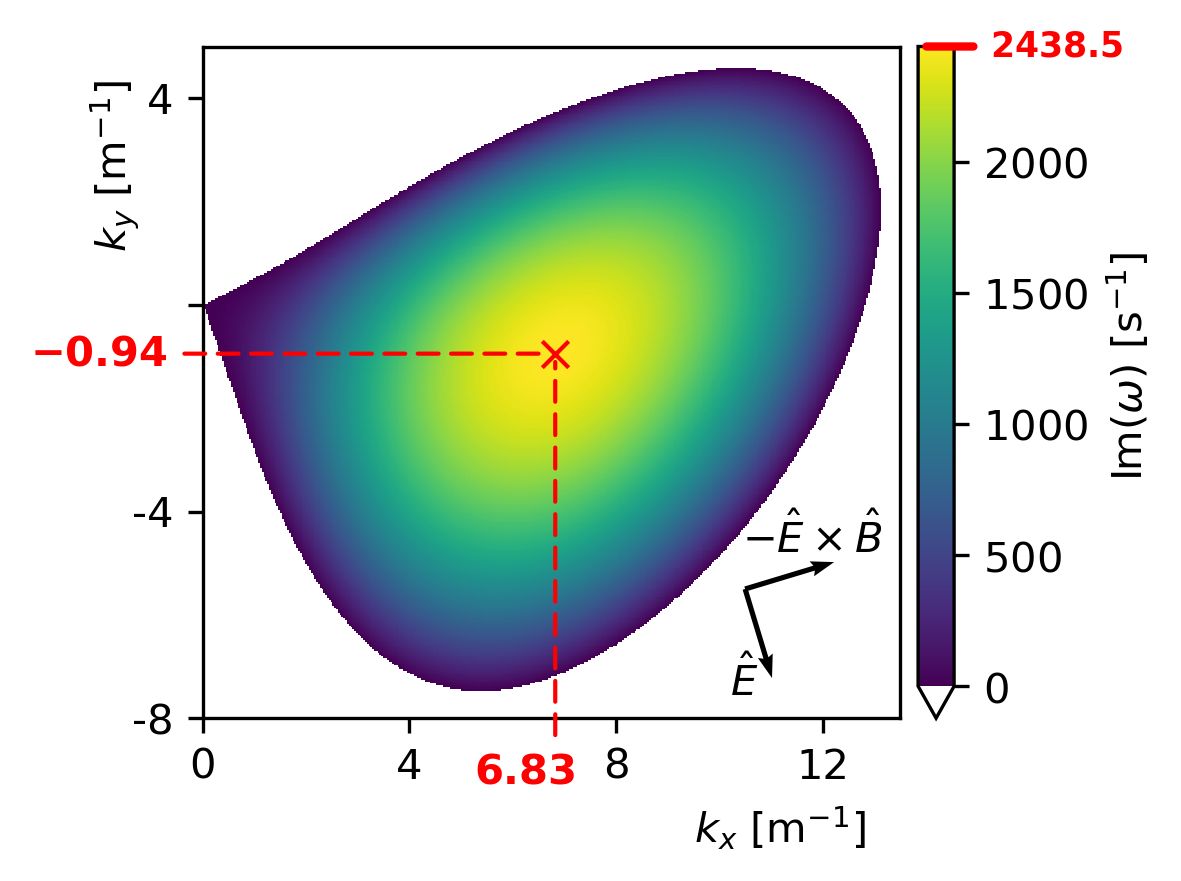}
      \caption{
        \label{fig:growthrate vs k theory}
        Predicted growth rate ($\text{Im}(\W)$) as a function of wavevector ($\K$), using $k_z = 0$, $\vec{B} = |\vec{B}| \hat{z}$, for the parameters shown in Table~\ref{tab:simulation params}. The black arrows in the lower right corner indicate the directions of $\vec{E}$ and $-\vec{E} \times \vec{B}$ for these conditions. The red annotations highlight the point with the maximum growth rate,
        indicating its value and location.
      }

    \end{figure}

    The direction of $\K$ gives insight into the physical mechanisms causing the instability. The thermal effects in the instability cause the largest growth rates for wavevectors parallel (or anti-parallel) to the bisector of $-\vec{E}$ and $\vec{E} \times \vec{B}$. Meanwhile the pure Farley-Buneman instability has maximum growth for wavevectors parallel (or anti-parallel) to the $\vec{E} \times \vec{B}$ direction \citep{Dimant2004, Dimant2022.TFBI}. 
    The wavevector at peak growth rate $\K_{\text{peak}}$ according to theory is $20\degr$ counterclockwise from the bisector of $\vec{E}$ and $-\vec{E} \times \vec{B}$, and $25\degr$ clockwise from the $-\vec{E} \times \vec{B}$ direction. This implies that for the chosen parameters, the thermal and the Farley-Buneman effects both play an important role.

    The length and time scales also help contextualize this instability. The wavelength at peak growth rate, 0.91~$m$, is much larger than the debye lengths ($\Ldebye{\A} =$ 90, 200, and 100~$\mu$m for H+, C+, and e-, respectively) and larger than the collisional mean free paths ($(\sqrt{T_\A/m_\A}) / \nu_{\A, H} = $ 0.02, 0.1, and 0.009~m for H+, C+, and e-, respectively).
    Meanwhile, the peak growth rate, $\text{Im}(\W_{\text{peak}}) = 2.4\times 10^{3}$~$s^{-1}$, and the wave frequency at that peak, $\text{Re}(\W_{\text{peak}}) = 2.0\times 10^{4}$~$s^{-1}$, correspond to timescales much smaller than those relevant to the macro-scale dynamics in the chromosphere \citep[see e.g.][]{Wedemeyer2004, Pereira2013, Carlsson2019}. 

    Finally, we gain further insight into this instability by considering the role of each ion species individually. Recalculating the theory using similar densities and temperatures but using H+ as the only ion species leads to smaller growth rate predictions with peak closer to $\K=0$. Repeating the calculation once more but this time using C+ as the only ion species leads to larger predicted growth rates which peak at larger $\K$. We conclude that both ions are important to the instability, with C+ driving the instability and H+ suppressing it.

  \subsection{Simulation Structure and Initial Conditions} \label{sec:model}

    To study a simplified case of this instability, we restrict ourselves to a 2D simulation using periodic boundary conditions including only electrons (e-), hydrogen neutrals (H), hydrogen ions (H+), and singly-ionized carbon (C+). We choose parameters, summarized in Table~\ref{tab:simulation params}, inspired by a cold region in the chromosphere where linear theory predicts the \TFBI\ will grow. We include singly-ionized Carbon in particular because initial studies of the \TFBIabbrv\ using PIC simulations and theory indicate that it is among the most important ionized species in determining the instability properties under chromospheric conditions \citep{Oppenheim2020}.

    \begin{table}   
      \centering
        \renewcommand{\arraystretch}{1.3}  
        \begin{tabular}{lrr|rrr|}
          \cline{2-6}
          \multicolumn{1}{l|}{} &
            \multicolumn{1}{c|}{$n$ {[}m$^{-3}${]}} &
            \multicolumn{1}{c|}{$u_x$ {[}m/s{]}} &
            \multicolumn{1}{c|}{$u_y$ {[}m/s{]}} &
            \multicolumn{1}{c|}{$T$ {[}K{]}} &
            \multicolumn{1}{c|}{$\nu_{\A, H}$ {[}s$^{-1}${]}} \\ \hline
          \multicolumn{1}{|l|}{e-} & \multicolumn{1}{r|}{$3.6 \times 10^{15}$} & - 8690 & \multicolumn{1}{r|}{- 1790} & \multicolumn{1}{r|}{7160} & $1.6 \times 10^{7}$      \\ \hline
          \multicolumn{1}{|l|}{C+} & \multicolumn{1}{r|}{$6.0 \times 10^{14}$} & - 1090 & \multicolumn{1}{r|}{- 4410} & \multicolumn{1}{r|}{4830} & $1.4 \times 10^{4}$      \\ \hline
          \multicolumn{1}{|l|}{H+} & \multicolumn{1}{r|}{$3.0 \times 10^{15}$} & + 190  & \multicolumn{1}{r|}{- 1260} & \multicolumn{1}{r|}{4060} & $6.7 \times 10^{5}$      \\ \hline
          \multicolumn{1}{|l|}{H}  & \multicolumn{1}{r|}{$8.0 \times 10^{19}$} & 0      & \multicolumn{1}{r|}{0}      & \multicolumn{1}{r|}{4000} & \multicolumn{1}{c|}{---} \\ \hline
          \hline  
          \multicolumn{3}{|c|}{$\vec{B}_{sim} = (10 \ \text{G}) \ \hat{z}$}             & \multicolumn{3}{c|}{$J_x^{(imposed)} = 5$ A/m$^2$}                                 \\ \hline
          \multicolumn{3}{|c|}{$\Delta_x=\Delta_y=2.5$ cm}                                          & \multicolumn{3}{c|}{$(N_x, N_y, N_z) = (512, 512, 1)$}                             \\ \hline
        \end{tabular}
      \caption{
        \label{tab:simulation params}
        Initial mean values of simulation parameters, representing a relatively cold region in the chromosphere. The table shows the means of number density ($n$), x- and y-components of velocity ($u_x$ and $u_y$), temperature ($T$), and momentum transfer collision frequency with neutrals ($\nu_{\A,H}$), for each fluid. The table also indicates the mean magnetic field in the simulation plane ($\vec{B}_{sim}$); the imposed current ($J_x^{(imposed)}$) as described by equations~\eqref{eqs:imposedcurrent}; the grid cell width in $x$ and $y$ ($\Delta_x$ and $\Delta_y$); and the number of cells in the $x$, $y$, and $z$ dimensions ($N_x$, $N_y$, and $N_z$).  
      }
    \end{table}


    In Table~\ref{tab:simulation params}, the mean values for ion densities, magnetic field, and neutral density, velocity, and temperature were chosen to represent a relatively cold region from a 2.5D radiative single-fluid MHD simulation of the chromosphere (see Section~\ref{sec:theory results}). The mean electron density satisfies quasineutrality \eqref{eq:QN}. The other initial mean velocities and temperatures are selected numerically such that the mean accelerations ($\partial \vec{u}_s / \partial t$) and temperature variations ($\partial T_s / \partial t$) of all other fluids are as close to zero as possible. These velocity and temperature selections bring the simulation conditions closer to the physics described by the theory, which assumes constant mean values. The electric field is determined by the electron momentum equation, assuming negligible electron interia; initially $\vec{E}(t{=}0) = (2.58\hat{x} - 8.53\hat{y})$~V/m, although later $\vec{E}$ changes as shown in Appendix~\ref{sec:E vs t}.

    The momentum transfer collision frequencies are calculated self-consistently, following the formalism of \citet[][and references therein]{Wargnier2022}. In particular, the (H+, H) collisions take into account the charge exchange resonance, and are treated as non-maxwellian. The (C+, H) collisions are treated assuming maxwell molecules. The (e-, H) collision frequency is calculated by performing the collision integral over experimentally determined differential cross sections. Coulomb collision frequencies would be orders of magnitude smaller than the other collision frequencies due to the small ionization fraction, however Coulomb collisions were instead turned off to simplify comparison between this simulation and the linear theory.


    The \TFBIabbrv\ must be driven by some energy source in order to grow. Given the chromospheric conditions selected in Table~\ref{tab:simulation params}, for a 2.5D multi-fluid simulation a sufficient source of energy can come from a current flowing across the box. Such a current can be caused by magnetic field lines bending out of the plane:
    \begin{subequations} \label{eqs:imposedcurrent}
      \begin{equation}  \label{eq:Bcurved}
        \vec{B} = \vec{B}_{sim} - z \mu_0 J_x^{(imposed)} \hat{y}
      \end{equation}
      \noindent where the simulation box is in the $xy$ plane, $z{=}0$; $\vec{B}_{sim} = \vec{B}(z{=}0)$ is the magnetic field in the simulation; and $J_x^{(imposed)}$ is some arbitrary value that determines the magnetic field line curvature. Bending the field lines affects the simulation only through spatial derivatives in $\vec{B}$, which only appear in the Ebysus model through Ampere's law \eqref{eq:curlB}. Plugging equation \eqref{eq:Bcurved} into Ampere's law yields:
      \begin{equation}  \label{eq:Jimposed}
        \mu_0 \vec{J} = \curl{\vec{B}_{sim}} + \mu_0 J_x^{(imposed)} \hat{x}
      \end{equation}
    \end{subequations}

    In our simulation, $\vec{B}_{sim}$ is constant except for a small spatial perturbation, with perturbation strength (the ratio between standard deviation and mean) peaking at $2.2\times 10^{-3}$, and always remaining less than 1.1\% of the electron density perturbation strength. Because the mean of $\vec{B}_{sim}$ is constant in time, the imposed current term provides the mean value for the current. For a nonzero current, the relative velocity differences between fluids enables energy transfer through collisions with neutrals, which may be sufficient to drive the \TFBIabbrv\ depending on the value of the current.

    We chose a current of 5~A/m$^2$ for the multi-fluid simulation, to reduce computational costs. This current is roughly 10 times larger than any currents found in the macro-scale simulated chromosphere discussed in Section~\ref{sec:theory results}. However, changing the current does not affect the linear theory of the \TFBIabbrv\ if all the ion densities also change by the same factor. Ion densities vary across many orders of magnitude in the simulated chromosphere, in some regions reaching at least 10 times smaller than those in Table~\ref{tab:simulation params}. Thus, the linear regime of the multi-fluid \TFBIabbrv\ simulation here is relevant to those regions in the chromosphere with the same ratios of current and ion densities, and the same values for other parameters.

    While the imposed current is sufficient to drive the \TFBIabbrv\ for our simulation, in theory it may be unnecessary. The imposed current serves to create sustained relative drifts between charged fluids and neutrals. Such sustained drifts might be generated without an imposed current in a significantly different parameter regime or with different fluids. However, for our simulation, removing the imposed current causes any velocity differences to vanish significantly faster than the instability growth rate.

  \subsection{Numerical Scheme} \label{sec:numerics}
    Ebysus \citep{MartinezSykora2020.Ebysus} is a multi-fluid radiative electromagnetic simulator designed to model the Sun's chromosphere. Here, we describe only the parts of the code used in our study of the \TFBI. For example, the Ebysus simulations here only utilize explicit methods, so we do not discuss the operator splitting option for semi-implicit time evolution. Some of the architecture and methodology in Ebysus are inherited from Bifrost \citep{Gudiksen2011}.
    
    Ebysus utilizes a 3rd-order predictor-corrector Hyman explicit timestep method \citep{Hyman1979} to calculate derivatives with respect to time.
    The numerical domain is defined in a staggered mesh, where values sometimes must be aligned in space. As necessary, interpolation is performed using a 5th-order scheme. Meanwhile, spatial derivatives are computed using a 6th-order scheme. The details of the staggered mesh, interpolation, and derivative calculations match those of Bifrost.

    Ebysus also includes artificial hyperdiffusion terms for stability, which primarily diffuse sharp fluctuations at small scales (5 grid cells or less). These terms are similar to those in Bifrost, but have been adapted to the multi-fluid model. Their exact forms are detailed in Appendix~\ref{sec:hyperdiffusion}.

\section{Results} \label{sec:results}
  Section~\ref{sec:theory results} discusses the predicted growth rate for the \TFBI\ throughout the chromosphere. This result comes from applying the linear theory to a single-fluid macro-scale simulation, and predicts that the instability occurs throughout many of the relatively cold regions in the chromosphere. Section~\ref{sec:simulation results} presents the main multi-fluid simulation in this work, and analyzes the simulation growth rates to confirm they match closely with theory. Section~\ref{sec:heating} demonstrates that this instability leads to increased temperatures and fluctuations in speed, as well as varied mean velocities, for all fluids in the simulation. Taken together, these results indicate that the effects of the \TFBIabbrv\ may significantly affect heating, transport, and random motions throughout the colder regions in the chromosphere.

  \subsection{Predicting Regions of Instability in the Chromosphere} \label{sec:theory results}
    To predict where the \TFBI\ occurs throughout the chromosphere, we combine the linear instability theory with output from a single-fluid macro-scale simulation run using the radiative MHD code, Bifrost \citep{Gudiksen2011}. This single-fluid simulation treats the hydrogen and helium ionization and recombination in non-equilibrium \citep{Leenaarts2007, Golding2016}, and incorporates some of the effects of interactions between ions and neutrals by including ambipolar diffusion \citep{NobregaSiverio2020}. Our prediction improves upon the related prediction in \citet{Oppenheim2020}, by solving the full multi-fluid linear theory including thermal perturbations, and utilizing output from a Bifrost simulation which included non-equilibrium-ionization modeling. 

    For this work, we convert the single-fluid Bifrost simulation output into a set of multi-fluid parameters including only H, H+, C+, and electrons. The magnetic field, along with the H and H+ density, come directly from the Bifrost simulation output, as the densities were tracked via the non-equilibrium-ionization modeling. The temperatures of all fluids are set equal to the simulated single-fluid temperature, for simplicity. The neutral velocity is set to zero, while the velocities of charged fluids come from the ambipolar velocity (Hall drift), as detailed in \citet{MartinezSykora2012}. Finally, the C+ density is set to the appropriate fraction of the single-fluid density, assuming photospheric abundances to find the density of carbon \citep{Asplund2009} and assuming statistical equilibrium to determine its ionization fraction.

\begin{figure*}  
      \includegraphics[width=\textwidth]{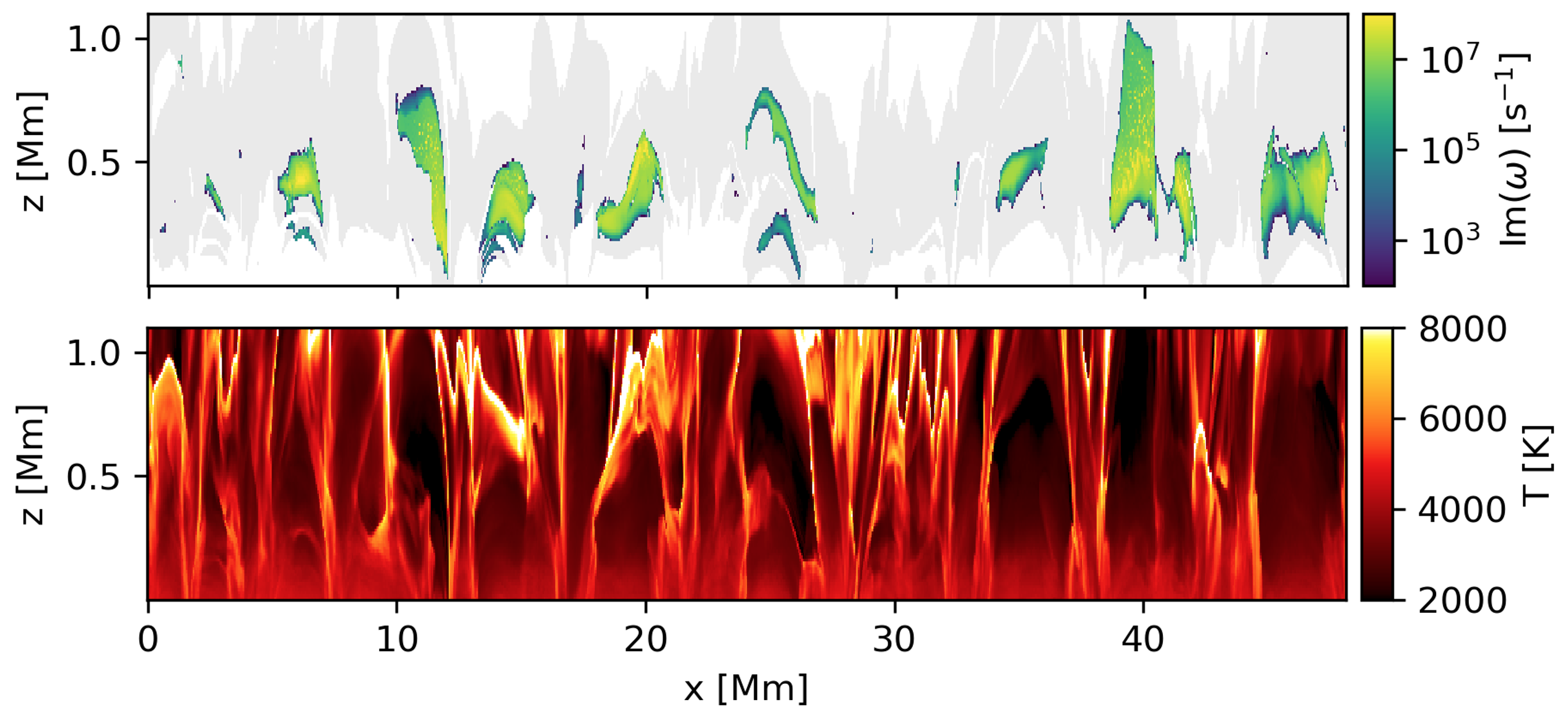}
      \caption{
        \label{fig:growthrate vs space}
        (Top) growth rate of the multi-fluid \TFBI\ throughout the chromosphere from a Bifrost simulation snapshot. The gray regions represent tested points with negative growth rates. The white regions show untested points where the assumptions of the \TFBIabbrv\ theory are not satisfied. Note that $x$ spans 50~Mm, while the $z$ direction ranges from 0 at the photosphere, up to 1.1~Mm. (Bottom) temperatures from the same Bifrost simulation snapshot. Many of the colder regions overlap with locations of predicted growth of the \TFBIabbrv.
      }
    \end{figure*}

    Figure~\ref{fig:growthrate vs space} shows the resulting growth rate prediction for the \TFBIabbrv\ throughout the simulated chromosphere. At each point in space, the growth rate is determined by taking the largest imaginary part of all the solutions for $\W$ across a variety of $\K$.
      We tested all values of $\K$ with magnitude 0.1, 0.3, 1, 3, 10, 30, 100, or 300 [$m^{-1}$],
      and each of 18 directions separated by 10 degree increments in the plane perpendicular to the local magnetic field.
    Points with negative growth rate are shown in gray.
    Note in particular that the predicted instability growth is correlated with the colder temperatures in the chromosphere.

    At every location with predicted growth, the single-fluid MHD model may be innacurate as it fails to incorporate the effects of the \TFBIabbrv. Combined with the prediction of heating due to the \TFBIabbrv\ (see Figure~\ref{fig:heating} in the next section), this supports the possibility of the \TFBIabbrv\ being responsible for the missing heating in chromospheric models.

    White regions in Figure~\ref{fig:growthrate vs space} indicate where the assumptions of the \TFBIabbrv\ theory break down, and the growth rate was not calculated. In the upper chromosphere and above, white regions indicate areas where the plasma does not satisfy the weakly ionized assumption, defined here as $n_{ion} / n_{neutral} < 0.01$.
    In the lower chromosphere and below, white regions indicate areas where the electrons are not strongly magnetized, having $|q_e| |\vec{B}| / (m_e \nu_{e, H}) < 2$. In regions of weakly magnetized or demagnetized electrons, we discovered that the \TFBIabbrv\ theory sometimes predicts instability growth (not shown on the plot), however it is only for large wavelengths ($|\K| \lesssim 0.01$~m$^{-1}$) and long timescales ($\text{Im}(\W) \lesssim 0.001$~s$^{-1}$). We mask these results because the lower solar atmosphere may be dynamic on such timescales \citep[see e.g.][]{Wedemeyer2004, Pereira2013, Carlsson2019} which invalidates the assumption of constant background as required by the linear theory. Furthermore, any physical mechanisms responsible for instability involving demagnetized electrons may be different than those responsible for the \TFBIabbrv.

    While Figure~\ref{fig:growthrate vs space} clearly suggests that the \TFBIabbrv\ occurs ubiquitously throughout the colder regions in the chromosphere, there are a few causes for concern about whether the numerical values of the predicted growth rates are similar to those in the actual chromosphere. Firstly, the underlying Bifrost simulation does not correctly represent the physics of \TFBIabbrv, and incorporating such effects may produce different results. In particular, large electric fields develop in Bifrost that indicate hypersonic drifts. These probably would be mitigated by the \TFBIabbrv. The instability would also cause heating and changes in velocity. Secondly, the assumptions that we applied to convert the single-fluid Bifrost simulation output into a set of multi-fluid values for the \TFBIabbrv\ theory could make these predictions inaccurate. Finally, gradients (e.g. in number density or temperature) are not included in the theory presented here, which assumes a constant background, but such gradients may affect the wave properties and growth rates.

\begin{figure*}  
      \includegraphics[width=\textwidth]{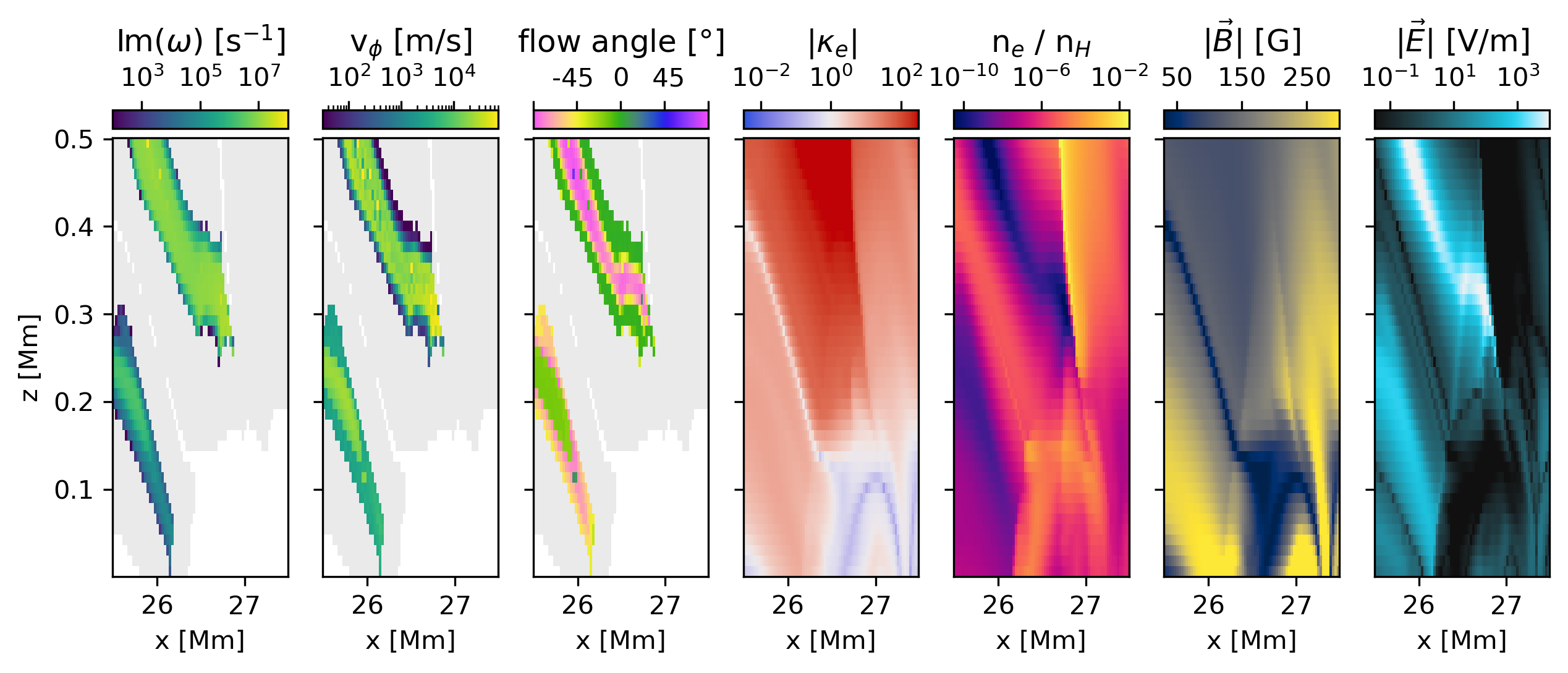}
      \caption{
        \label{fig:multiplot}
        Various parameters plotted in the region $25.5$~Mm~$<x<27.5$~Mm, $0<z<0.5$~Mm using the same Bifrost snapshot as in Figure~\ref{fig:growthrate vs space}. From left to right, these panels show: the predicted growth rates, when testing more values of $\K$ than those used in Figure~\ref{fig:growthrate vs space}; the phase speed, $v_{\phi} \eqdef \text{Re}(\W) / |\K|$; the flow angle, i.e. the angle from $\vec{E} \times \vec{B}$ to $\K$ or $-\K$, whichever is closer; the electron magnetization, $|\kappa_e| = |q_e| |\vec{B}| / (m_e \nu_{e,H})$; the ionization fraction, $n_e / n_H$; the magnetic field magnitude, $|\vec{B}|$; and the electric field magnitude, $|\vec{E}|$. In the first three panels, gray indicates negative growth rate predictions, while white corresponds with untested points where the \TFBIabbrv\ assumptions break down.
      }
    \end{figure*}

    Due to the limitations of this analysis and the Bifrost model, we further explore these predictions in Figure~\ref{fig:multiplot}. This figure shows various parameters in one particular area where there are two distinct regions of predicted instability growth. First, we check that the previous prediction was not missing any significant regions of instability, by sweeping across more possible values of $\K$. The leftmost panel of Figure~\ref{fig:multiplot} shows the predicted growth rates
      after checking values of $\K$ with each of 24 magnitudes between 0.1 and 681 [$m^{-1}$] (inclusive) with even logarithmic spacing,
      and each of 60 directions separated by 3 degree increments in the plane perpendicular to the local magnetic field.
    This more accurate search predicts that the instability will occur in the same regions as in Figure~\ref{fig:growthrate vs space}, though with slightly larger growth rates.

    The second and third panels in Figure~\ref{fig:multiplot} further characterize the predicted wave properties in this region. The second panel shows the phase speed, $v_{\phi} \eqdef \text{Re}(\W) / |\K|$. The third panel shows the flow angle, the angle from $\vec{E} \times \vec{B}$ to $\K$ or $-\K$, whichever is closer. This angle gives insight into which effects contribute to the instability. In the single-species ion case with strongly magnetized electrons and weakly magnetized ions, pure Farley-Buneman waves have a flow angle near $0\degr$, while waves dominated by thermal effects have a flow angle close to $-45\degr$, the bisector of $-\vec{E}$ and $\vec{E} \times \vec{B}$ \citep{Dimant2004, Dimant2022.TFBI}. Considering the lower left area of predicted instability, this implies that the instabilities near the edges of this area may be dominated mainly by thermal effects, while the instabilities near its center may have significant contributions from both thermal and Farley-Buneman effects --- requiring the \TFBI\ theory for an accurate description.

    The remaining panels in Figure~\ref{fig:multiplot} provide some other physical parameters for reference. The fourth panel provides the electron magnetization, $|\kappa_e| = |q_e| |\vec{B}| / (m_e \nu_{e,H})$, which is larger than 2.0 for all non-white points in the first three panels. The fifth panel shows the ionization fraction, $n_e / n_H$, which is smaller than 0.01 for all non-white points in the first three panels. The sixth panel plots the magnitude of the magnetic field, $|\vec{B}|$. The final panel shows the magnitude of the electric field, $|\vec{E}|$, which reaches to more than 1000 V/m in some regions; such large electric fields could be mitigated by the presence of the \TFBIabbrv, which is not incorporated into the Bifrost simulation. Note in particular that the areas of predicted growth for the \TFBIabbrv\ are dictated by the physical parameters, which form into complicated shapes rather than follow any sort of simple layering scheme in the chromosphere.

  \subsection{Simulation of the Instability} \label{sec:simulation results}
    We use Ebysus to run a multi-fluid simulation of a relatively cold region in the chromosphere with magnetic field lines bent out of the plane, using the parameters in Table~\ref{tab:simulation params}. This simulation shows a clear wave pattern similar to that found in kinetic simulations \citep{Oppenheim2020}. The growth rate agrees with linear theory during the linear regime, indicating an accurate reproduction of the \TFBI. This success demonstrates that multi-fluid simulators are capable of producing the \TFBIabbrv.

\begin{figure*}  
      \includegraphics[width=\textwidth]{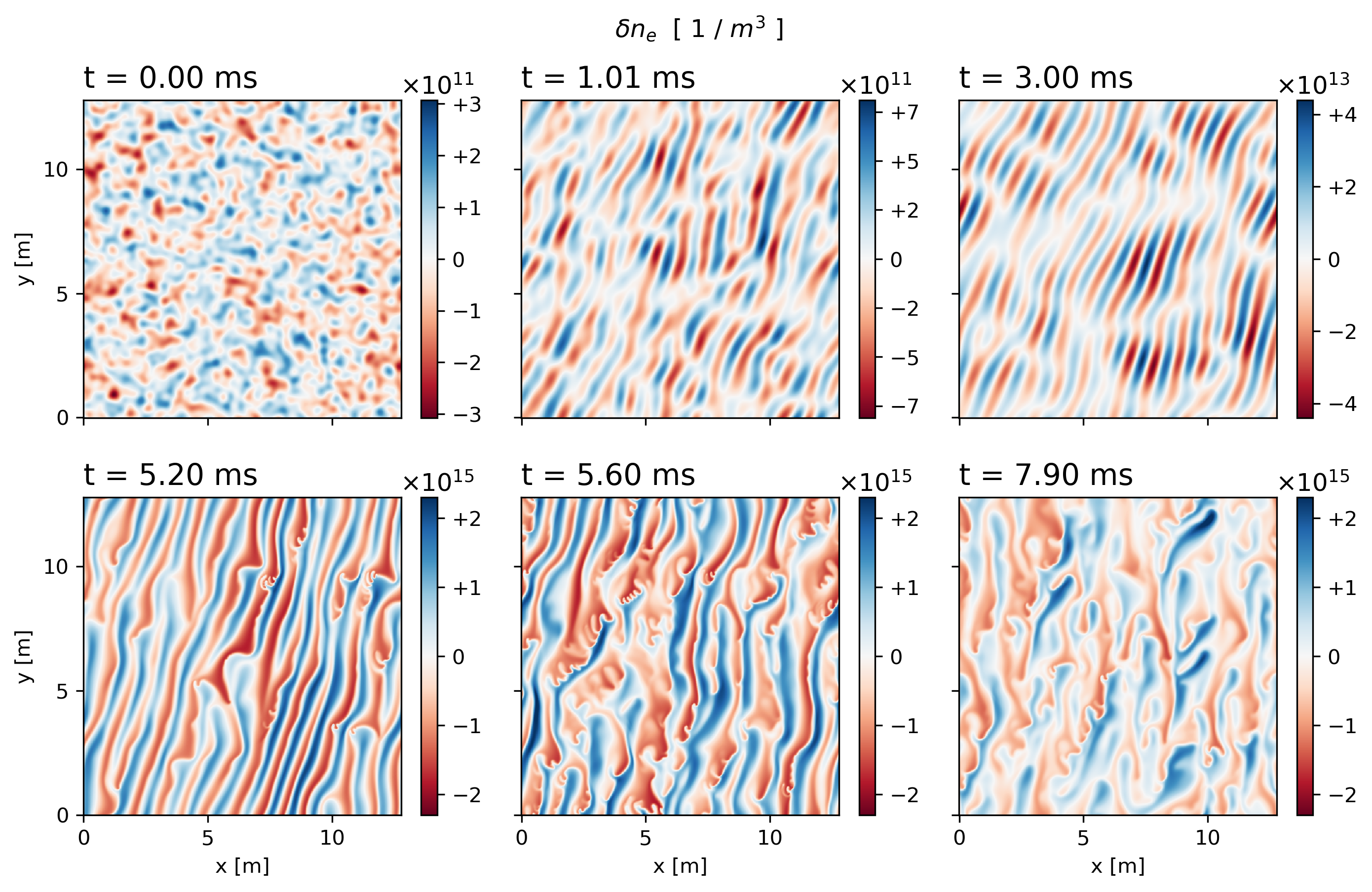}
      \caption{
        \label{fig:simulation}
        Perturbation of electron number density at selected snapshots throughout the simulation. The associated movie available online shows the evolution throughout the entire simulation. Note that the color scale varies between panels in this figure, and varies in time in the animation. Panels show snapshots of the simulation at different stages, including the initial conditions, the linear growth phase, and the nonlinear regime.
      }
    \end{figure*}

    Figure~\ref{fig:simulation} and the corresponding animation show the electron number density throughout the simulation. We initialize the number density at $t=0$ (top left panel) with a random perturbation having standard deviation approximately 4.6 orders of magnitude smaller than the background density, smoothed by a gaussian kernel to mitigate numerical artifacts at the grid scale. A clear wave pattern develops by $t=1.0$~ms (top middle panel), and the perturbation grows according to linear theory. By $t=3.0$~ms (top right panel), the perturbation has grown by roughly two orders of magnitude. At around $t\approx5.2$~ms (bottom left panel), nonlinear effects start to develop, as the perturbation becomes comparable in magnitude to the mean density, $3.6 \times 10^{15}$~$m^{-3}$.

    Around $t\approx5.3$~ms, the root-mean-square perturbation reaches its maximum of roughly 27\% of the background value, though it eventually settles down to roughly 17\% by the end of the simulation. The bottom middle panel of Figure~\ref{fig:simulation} shows that by $t=5.6$~ms, secondary waves have developed and spread throughout the simulation box. From this time onwards, the linear-stage \TFBIabbrv\ is no longer the dominant effect in the simulation. Finally, the bottom right panel shows the density when the simulation ends at $t=7.9$~ms. By the end of the simulation, the perturbations reach a quasi-steady state where the amplitude and scale size of features settle to roughly constant values.

\begin{figure*}  
      \includegraphics[width=\textwidth]{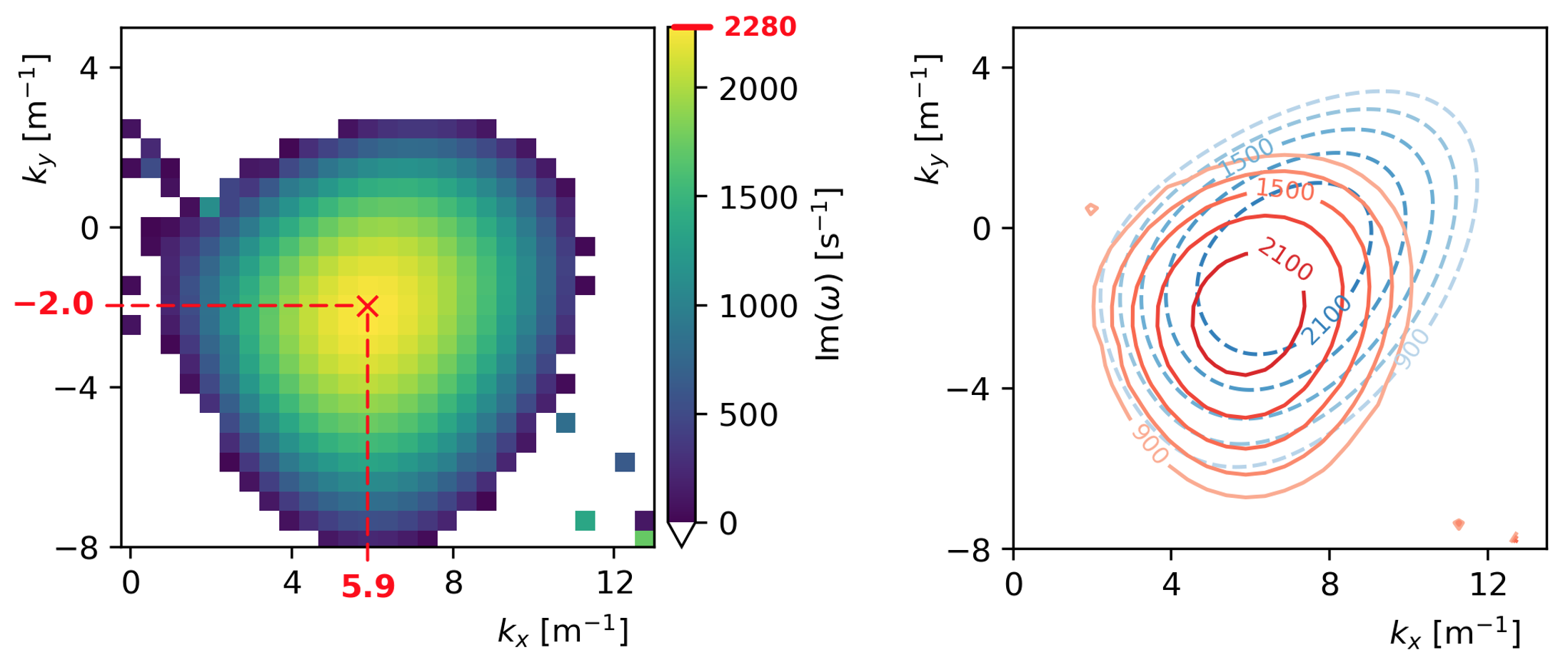}
      \caption{
        \label{fig:growthrate vs k simu compare}
        (Left) growth rate as a function of wavevector, calculated using simulation outputs during the linear growth stage. (Right) contours of the growth rate versus $\K$ map from the simulation (red, solid lines) and the theoretical prediction from Figure~\ref{fig:growthrate vs k theory} (blue, dashed lines), labeled with values in units of s$^{-1}$. The close agreement supports the claim that this simulation reproduces the \TFBI. The remaining discrepancy between theory and simulation is addressed further in the main text.
      }
    \end{figure*}

    To confirm that this simulation really does reproduce the \TFBIabbrv, Figure~\ref{fig:growthrate vs k simu compare} compares growth rates from the simulation to theory. To determine the growth rates, we compute a Fourier transform in space at each snapshot in time, $\mathcal{F}_t(\K)$, of the electron number density perturbation from $t=0.6$~ms to $t=1.9$~ms. According to linear theory, the magnitude at each $\K$ should scale as $\exp \brackets{\text{Im}(\W) t}$. Thus, for each $\K$, the slope of the best fit line through the natural log of the magnitude of the Fourier transforms provides the simulation growth rate, $\gamma \eqdef \text{Im}(\W)$, as follows:
    \begin{equation}
      \gamma \ t + \text{offset} = \text{ln}\enc{|}{ \mathcal{F}_t(\K) }{|}
    \end{equation}

    \noindent The left panel of Figure~\ref{fig:growthrate vs k simu compare} plots the results of this fitting process. The right panel of the figure compares simulation and theory directly by overlaying contours of $\gamma$ as determined here for the simulation, and in Section~\ref{sec:theory} for the theory.

    Figure~\ref{fig:growthrate vs k simu compare} shows remarkably close agreement between simulation and theory. Comparing qualitatively at the peak growth rates, the simulation growth rate $\W_{\text{peak}}^{\text{(sim)}}$ is $6.6\%$ less than $\W_{\text{peak}}$ in the theory. The magnitude of the wavevector at the simulation peak $|\K_{\text{peak}}^{\text{(sim)}}|$ is $9.6\%$ smaller than in the theory, and its direction $\K_{\text{peak}}^{\text{(sim)}}$ differs from the theory peak by $10.8\degr$. From this close agreement, we conclude that this simulation does indeed reproduce the \TFBIabbrv\ described by linear theory.

    A small discrepancy still remains between simulated and predicted growth rate versus wavevector distributions. One possible source of error is the changing background quantities. In particular, the theory neglects any zeroth order acceleration ($d_\A \vec{u}_\A/dt$), and temperature variations ($d_\A T_\A/dt$). Meanwhile, some background acceleration and heating in the simulation is an unavoidable consequence of the imposed current (see equations \eqref{eqs:imposedcurrent}), although the amount depends on the simulation parameters.

    To check whether the zeroth-order effects of imposed current are the main source of the discrepancy, we repeated the simulation but using imposed current and ion number densities 10 times larger (not shown here). This change of parameters has almost no effect on the theoretical prediction, while increasing the zeroth-order acceleration and heating of all electrons and ions by a factor of 10. The discrepancy between simulation and theory also increases significantly. Quantitatively, at the peak for this test simulation, the growth rate is $11.6\%$ less than in the theory (compare with $6.6\%$ from the main simulation), $|\K|$ is $7.4\%$ smaller than in the theory (compare with $9.6\%$), and the direction of $\K$ differs from the theory peak by $14.7\degr$ (compare with $10.8\degr$). We conclude that the zeroth-order acceleration and heating terms are the most likely main source of error in the original simulation.

    Other possible sources of error include electromagnetic effects, which are included in the simulation but not the theory, and any artifacts of the numerical method used for the simulation. The small magnetic field fluctuations, with relative size less than 1\% compared to the relative size of density fluctuations, suggest the electromagnetic assumption does not introduce a sizeable error. Meanwhile, we found the numerical diffusion effects to be small, especially during the linear growth stages of the simulation, implying at most minor error contributions from numerical artifacts.

    The linear stage of the main \TFBIabbrv\ simulation confirms the instability occurs for the chromospheric parameters in Table~\ref{tab:simulation params}, as well as for any similar plasma with the same ratios of current and ion densities. In particular, the simulation also reproduces the linear stage of the \TFBIabbrv\ for such plasma, because the linear theory is unaffected by changing current and ion densities by the same factor. Additionally, the trend from test simulation to main simulation suggests that similar simulations with even smaller current and ion densities would have even better agreement between the linear regime of the simulation and the linear theory.

    In the next section, we analyze the effects of turbulence throughout this simulation. While we are confident that this simulation accurately represents the linear regime for any similar plasma with the same ratios of current and ion densities, it is not yet clear how the nonlinear behavior would be altered by using different parameters. 


  \subsection{Effects of Turbulence --- Heating, Transport, and Random Motion} \label{sec:heating}
    While the linear theory fully breaks down at around $t=5.2$~ms in our simulation of the \TFBI, turbulence affects the temperatures and velocities of fluids as soon as $t=4.0$~ms, when the r.m.s. electron density perturbation reaches approximately $3\%$ of the mean electron density.
    These non-linear effects do not arise physically in macro-scale models which fail to resolve the small-scales (a few meters, and a few milliseconds) and to include the multi-fluid physics relevant to the \TFBIabbrv. Therefore, effects of the \TFBIabbrv\ might cause disagreements when comparing such models to solar observations. In this section, we use our simulation to investigate the turbulence-driven heating, transport, and random motions due to the \TFBIabbrv.

\begin{figure}  
      \includegraphics[width=\customhalfwidth]{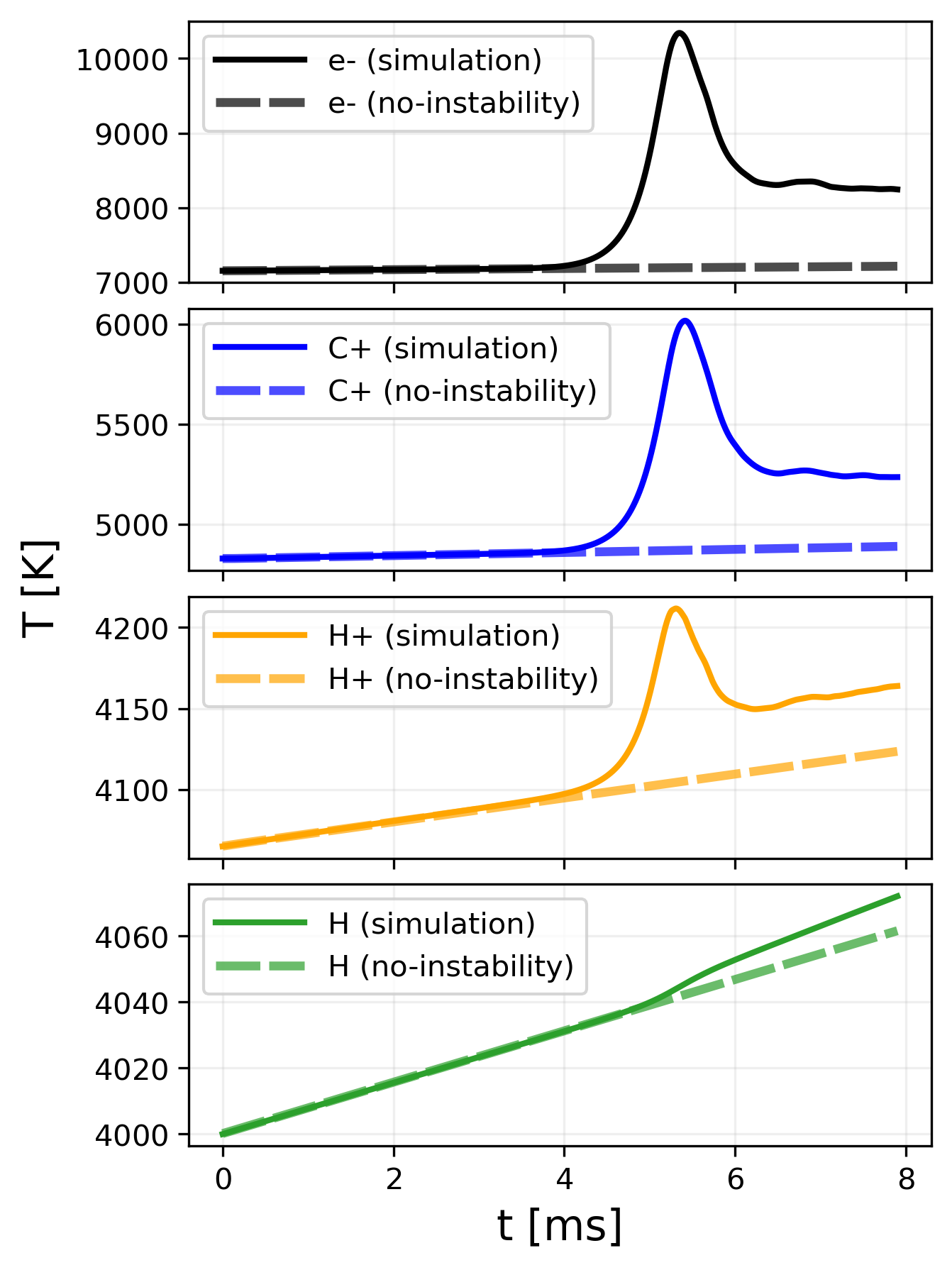}
      \caption{
        \label{fig:heating}
        Temperature evolutions of each fluid throughout the simulation of the \TFBIabbrv. Solid lines show mean temperatures throughout the simulation. Dashed lines show temperature predictions using the same physical parameters but without accounting for the instability. Linear theory alone predicts no change in mean temperatures due to the instability; around $t=4.0$~ms, nonlinear effects start to become important and cause heating. 
      }
    \end{figure}

    Figure~\ref{fig:heating} illustrates the turbulence-driven heating in the simulation. The plots show the evolutions of fluids' temperatures throughout the simulation, as well as the temperature evolution predictions for a no-instability model with the same physical parameters as in the simulation (shown in Table~\ref{tab:simulation params}) but which lacks the spatial resolution to reproduce the \TFBIabbrv. These no-instability temperature predictions are constructed by plugging mean values of quantities into the energy equation \eqref{eq:heating} to calculate $\partial T_\A/ \partial t$ from $t=0$ to $t=0.5$~ms --- when the instability effects become relevant --- then extrapolating linearly until the end of the simulation. The no-instability model shows constant nonzero heating due to the imposed current (see Eqs \ref{eqs:imposedcurrent}).

    There is significant heating due to the \TFBI. In the simulation, the electron temperature overshoots up to 3000~K more than its original value of 7000~K, before settling down to about 8300~K, 1300~K above the original temperature. The ion temperatures look qualitatively similar: C+ peaks at an increase of 1200~K before settling to 400~K above the no-instability model temperature, while H+ peaks at an increase of 100~K and settles to an increase of 30~K. The neutral temperature does not overshoot, but ends up approximately 10~K larger by the end of the simulation due to thermalization with the other fluids that all heat up in the presence of the \TFBIabbrv. This heating may contribute towards heating the actual chromosphere, and may help explain why macro-scale models such as Bifrost predict temperatures thousands of Kelvin smaller than those implied by observations. 

    The heating comes from collisional effects. Collisions convert the kinetic energy into thermal energy, and allow fluids to thermalize with each other. The dissipation of velocity drifts heats the ions and electrons, though a majority of that thermal energy transfers into the neutrals. Still, the neutral temperature changes less than the other fluids' temperatures because the neutrals are multiple orders of magnitude denser.

\begin{figure*}  
      \includegraphics[width=\textwidth]{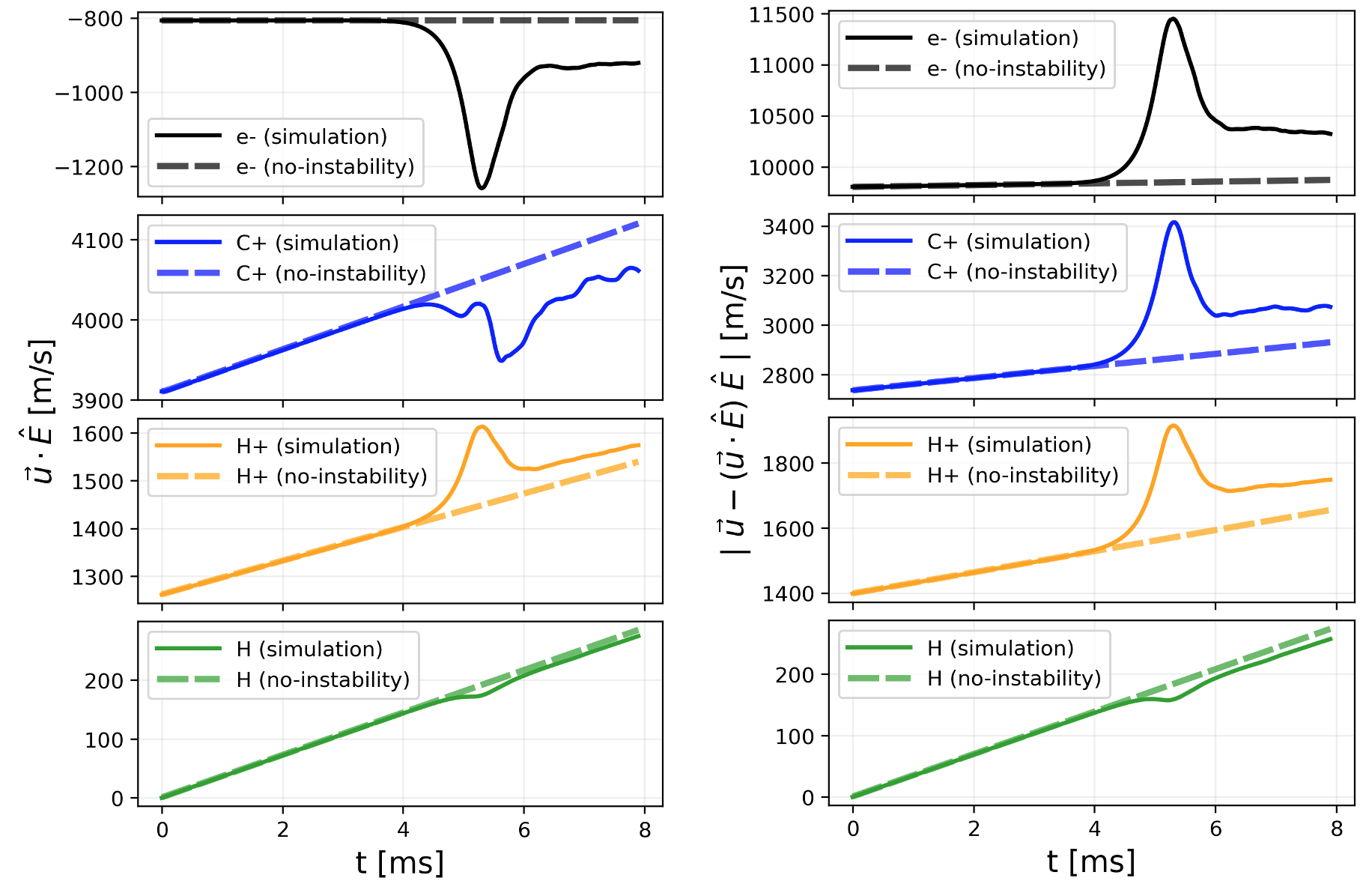}
      \caption{
        \label{fig:transport}
        Velocity evolutions throughout the \TFBIabbrv\ simulation, for each fluid (electrons, C+, H+, H from top to bottom). Solid lines show mean velocities throughout the simulation. Dashed lines show velocity predictions using the same physical parameters but not accounting for any instability. The left panels show the velocity component in the electric field direction, $\vec{u} \cdot \hat{E}$, while the right panels show the component perpendicular to the electric field, $|\vec{u} - (\vec{u} \cdot \hat{E}) \hat{E}|$, where $\hat{E} \eqdef \vec{E} / |\vec{E}|$. Linear theory alone predicts no change in mean velocities due to the instability; around $t=4.0$~ms, nonlinear effects start to become important and cause acceleration. 
      }
    \end{figure*}

    Figure~\ref{fig:transport} illustrates the turbulence-driven transport in the simulation. The plots show the evolutions of fluids' velocities throughout the simulation, broken up into components parallel and perpendicular to the mean electric field. Similarly to Figure~\ref{fig:heating}, these plots also compare to a no-instability model, constructed here by plugging mean values into the momentum equation \eqref{eq:momentum} to calculate the accelerations until $t=0.5$~ms, then extrapolating linearly after that time. The no-instability model has a nonzero slope due to the imposed current (see Eqs \ref{eqs:imposedcurrent}).

    There is moderate transport due to the \TFBI\ in this simulation. For each velocity component of each fluid, the non-linear effects are not apparent until roughly $t=4.0$~ms, at which point the behavior changes, leading to an overshoot then settling towards some particular deviation from the no-instability model. Parallel to $\vec{E}$, the electrons end up with a velocity of roughly -920~m/s, 120~m/s less than the no-instability model predicts. The ion and neutral velocities in this direction all differ from the no-instability model by less than 5\%. Perpendicular to $\vec{E}$, the electrons end up with a simulation mean velocity which is roughly 450~m/s (5\%) larger due to the instability. The ion velocities in this direction increase by roughly 5\% due to the instability, while the neutral velocity decreases by roughly 5\%.

    Altering the mean velocities affects the electric field strength and direction. Electrons travelling parallel to $\vec{E}$ work to short out the field, while those travelling perpendicular to $\vec{E}$ increase the field. For our simulation, the impact of increased perpendicular transport is stronger than the change in transport pallel to $\vec{E}$, leading to an increased electric field magnitude, as shown in Appendix~\ref{sec:E vs t}. To incorporate these effects into a macro-scale model, more work is required to determine the behavior of instability-driven transport and electric field changes across a range of parameters. Eventually, these effects could be modeled by parametrically adjusting electron and ion collision frequencies with neutrals, altering the effective cross-field conductivities.

\begin{figure}  
      \includegraphics[width=\customhalfwidth]{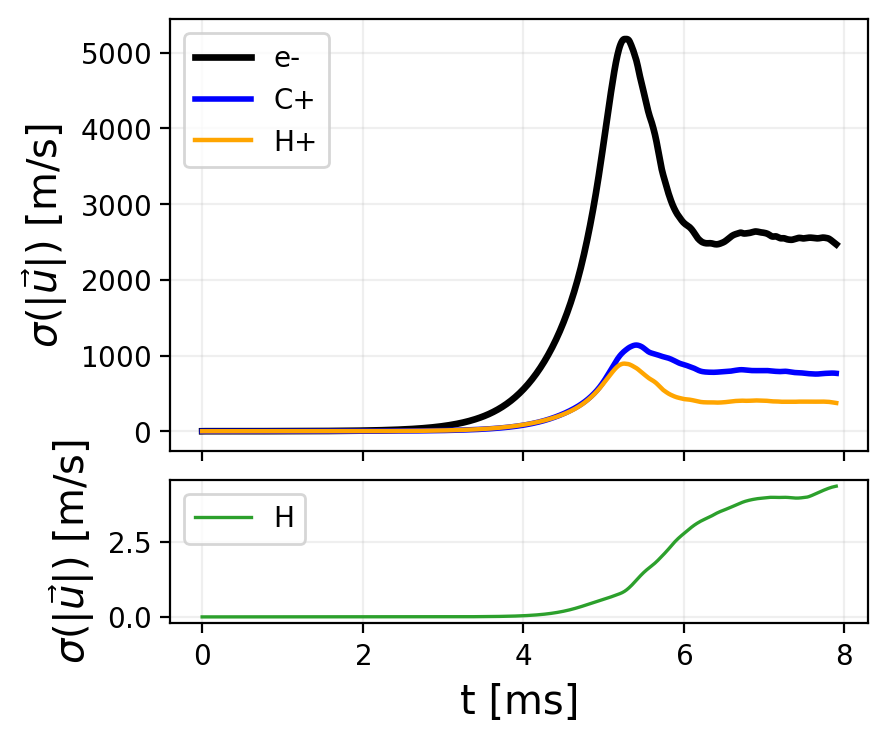}
      \caption{
        \label{fig:motion}
        Random motions of each fluid throughout the simulation of the \TFBIabbrv. The y-axis shows the standard deviation of the fluid speed, and is split into top and bottom plots with different scaling so that the changes for neutral hydrogen are visible. 
      }
    \end{figure}

    Figure \ref{fig:motion} shows the random motions of each fluid throughout the main multi-fluid simulation. These motions are computed by taking the standard deviation of the speed (i.e., magnitude of velocity) for each fluid at each simulation snapshot. Similarly to the turbulence-driven heating, the random motion speeds of all the charged fluids overshoot, then settle down to some value above a baseline. The relevant baseline in this case is 0; a model lacking the resolution to consider fluctuations would see zero deviation from the mean caused by effects at this scale.

    There are notable random motions due to the \TFBIabbrv. The standard deviation of electron speed overshoots to 5200~m/s before settling to roughly 2500~m/s. For C+, the peak is around 1150~m/s before settling to roughly 750~m/s. For H+, the peak is at 900~m/s, and the random motion speed decreases to 400~m/s by the end of the simulation. The neutral speed does not overshoot but ends up at approximately 4~m/s, due to collisions with the other fluids in the simulation. These random motions would contribute to broadening spectral lines in observations, and their values are consistent with the so-called ``microturbulence'' inferred through inversions of observations \citep{daSilvaSantos2020}.

\section{Conclusions} \label{sec:conclusion}
  Combining the linear instability theory of a multi-fluid model and the output of a single-fluid simulation, this work predicts that the \TFBI\ occurs throughout many of the colder regions in the chromosphere. This prediction improves upon the related prediction in \citet{Oppenheim2020} by solving the full multi-fluid linear theory including thermal perturbations, and utilizing output from a Bifrost simulation which included non-equilibrium-ionization modeling. Our estimates reveal that the single-fluid radiative MHD model has extended regions which may be innacurate since the model does not incorporate effects of the \TFBIabbrv. 

  Focusing on the parameters found in one of these colder regions in the chromosphere, we produce the first multi-fluid simulation of the \TFBIabbrv. We validate this by showing close agreement between the simulation and linear theory. For computational reasons, we used a current that is too large by roughly an order of magnitude, compared to those in the single-fluid simulated chromosphere. This adjustment does not affect the linear theory, but likely contributes to the small error between simulation and theory during the linear regime. The ability to produce this instability with a multi-fluid code enables further study of the instability across chromospheric parameter ranges which are computationally challenging for kinetic models.

  Our multi-fluid simulation exhibits turbulence-driven heating, transport, and enhanced random motions of all fluids in the simulation. The significant heating supports the possibility that the \TFBIabbrv\ may contribute towards chromospheric heating. The transport will modify cross-field conductivities and electric fields, and the random motions should broaden spectral lines in observations. However, non-linear effects may behave differently for different sets of parameters throughout the chromosphere. Determining quantitatively the impacts of the \TFBIabbrv\ throughout this complex region may require a suite of small-scale multi-fluid simulations spanning a wide range of parameters.

  The non-linear effects caused by the \TFBIabbrv\ occur on scales of meters and milliseconds --- far smaller than what has been resolved by macro-scale simulations of the Sun's atmosphere --- yet they may play an important role in explaining observations of chromospheric heating and line-broadening due to random motion. These effects motivate further study of the \TFBI\ and its impact throughout the chromosphere.

\begin{acknowledgments}
  \handleAckLineNumbers
  Acknowledgements: SE, MO, YD, and RX gratefully acknowledge the support of this work by NSF Grant No. 1903416. JMS gratefully acknowledges the support of NASA grants 80NSSC20K1272, 80NSSC21K0737, 80NSSC21K1684, and contract NNG09FA40C (\IRIS). Resources supporting this work were provided by the NASA High-End Computing (HEC) Program through the NASA Advanced Supercomputing (NAS) Division at Ames Research Center. The simulation has been run on Pleiades cluster through the computing projects s1061, s2601.
\end{acknowledgments}

\bibliography{bibfile}{}

\begin{thebibliography}{}
\expandafter\ifx\csname natexlab\endcsname\relax\def\natexlab#1{#1}\fi
\providecommand{\url}[1]{\href{#1}{#1}}
\providecommand{\dodoi}[1]{doi:~\href{http://doi.org/#1}{\nolinkurl{#1}}}
\providecommand{\doeprint}[1]{\href{http://ascl.net/#1}{\nolinkurl{http://ascl.net/#1}}}
\providecommand{\doarXiv}[1]{\href{https://arxiv.org/abs/#1}{\nolinkurl{https://arxiv.org/abs/#1}}}

\bibitem[{{Asplund} {et~al.}(2009){Asplund}, {Grevesse}, {Sauval}, \&
  {Scott}}]{Asplund2009}
{Asplund}, M., {Grevesse}, N., {Sauval}, A.~J., \& {Scott}, P. 2009, \araa, 47,
  481, \dodoi{10.1146/annurev.astro.46.060407.145222}

\bibitem[{{Ballester} {et~al.}(2018){Ballester}, {Alexeev}, {Collados},
  {Downes}, {Pfaff}, {Gilbert}, {Khodachenko}, {Khomenko}, {Shaikhislamov},
  {Soler}, {V{\'a}zquez-Semadeni}, \& {Zaqarashvili}}]{Ballester2018}
{Ballester}, J.~L., {Alexeev}, I., {Collados}, M., {et~al.} 2018, \ssr, 214,
  58, \dodoi{10.1007/s11214-018-0485-6}

\bibitem[{{Carlsson}(2007)}]{Carlsson2007}
{Carlsson}, M. 2007, in Astronomical Society of the Pacific Conference Series,
  Vol. 368, The Physics of Chromospheric Plasmas, ed. P.~{Heinzel},
  I.~{Dorotovi{\v{c}}}, \& R.~J. {Rutten}, 49.
\newblock \doarXiv{0704.1509}

\bibitem[{{Carlsson} {et~al.}(2019){Carlsson}, {De Pontieu}, \&
  {Hansteen}}]{Carlsson2019}
{Carlsson}, M., {De Pontieu}, B., \& {Hansteen}, V.~H. 2019, \araa, 57, 189,
  \dodoi{10.1146/annurev-astro-081817-052044}

\bibitem[{{Carlsson} \& {Stein}(1992)}]{Carlsson1992}
{Carlsson}, M., \& {Stein}, R.~F. 1992, \apjl, 397, L59, \dodoi{10.1086/186544}

\bibitem[{{Carlsson} \& {Stein}(1995)}]{Carlsson1995}
---. 1995, \apjl, 440, L29, \dodoi{10.1086/187753}

\bibitem[{{Carlsson} \& {Stein}(2002)}]{Carlsson2002}
---. 2002, \apj, 572, 626, \dodoi{10.1086/340293}

\bibitem[{{Cheung} \& {Isobe}(2014)}]{Cheung2014}
{Cheung}, M. C.~M., \& {Isobe}, H. 2014, Living Reviews in Solar Physics, 11,
  3, \dodoi{10.12942/lrsp-2014-3}

\bibitem[{{Chintzoglou} {et~al.}(2021){Chintzoglou}, {De Pontieu},
  {Mart{\'\i}nez-Sykora}, {Hansteen}, {de la Cruz Rodr{\'\i}guez},
  {Szydlarski}, {Jafarzadeh}, {Wedemeyer}, {Bastian}, \& {Sainz
  Dalda}}]{Chintzoglou2021}
{Chintzoglou}, G., {De Pontieu}, B., {Mart{\'\i}nez-Sykora}, J., {et~al.} 2021,
  \apj, 906, 82, \dodoi{10.3847/1538-4357/abc9b1}

\bibitem[{{da Silva Santos} {et~al.}(2020){da Silva Santos}, {de la Cruz
  Rodr{\'\i}guez}, {Leenaarts}, {Chintzoglou}, {De Pontieu}, {Wedemeyer}, \&
  {Szydlarski}}]{daSilvaSantos2020}
{da Silva Santos}, J.~M., {de la Cruz Rodr{\'\i}guez}, J., {Leenaarts}, J.,
  {et~al.} 2020, \aap, 634, A56, \dodoi{10.1051/0004-6361/201937117}

\bibitem[{{De Pontieu} {et~al.}(2014){De Pontieu}, {Title}, {Lemen}, {Kushner},
  {Akin}, {Allard}, {Berger}, {Boerner}, {Cheung}, {Chou}, {Drake}, {Duncan},
  {Freeland}, {Heyman}, {Hoffman}, {Hurlburt}, {Lindgren}, {Mathur}, {Rehse},
  {Sabolish}, {Seguin}, {Schrijver}, {Tarbell}, {W{\"u}lser}, {Wolfson},
  {Yanari}, {Mudge}, {Nguyen-Phuc}, {Timmons}, {van Bezooijen}, {Weingrod},
  {Brookner}, {Butcher}, {Dougherty}, {Eder}, {Knagenhjelm}, {Larsen},
  {Mansir}, {Phan}, {Boyle}, {Cheimets}, {DeLuca}, {Golub}, {Gates}, {Hertz},
  {McKillop}, {Park}, {Perry}, {Podgorski}, {Reeves}, {Saar}, {Testa}, {Tian},
  {Weber}, {Dunn}, {Eccles}, {Jaeggli}, {Kankelborg}, {Mashburn}, {Pust},
  {Springer}, {Carvalho}, {Kleint}, {Marmie}, {Mazmanian}, {Pereira}, {Sawyer},
  {Strong}, {Worden}, {Carlsson}, {Hansteen}, {Leenaarts}, {Wiesmann},
  {Aloise}, {Chu}, {Bush}, {Scherrer}, {Brekke}, {Martinez-Sykora}, {Lites},
  {McIntosh}, {Uitenbroek}, {Okamoto}, {Gummin}, {Auker}, {Jerram}, {Pool}, \&
  {Waltham}}]{DePontieu2014}
{De Pontieu}, B., {Title}, A.~M., {Lemen}, J.~R., {et~al.} 2014, \solphys, 289,
  2733, \dodoi{10.1007/s11207-014-0485-y}

\bibitem[{{Dimant} \& {Oppenheim}(2004)}]{Dimant2004}
{Dimant}, Y.~S., \& {Oppenheim}, M.~M. 2004, Journal of Atmospheric and
  Solar-Terrestrial Physics, 66, 1639, \dodoi{10.1016/j.jastp.2004.07.006}

\bibitem[{{Dimant} {et~al.}(2022){Dimant}, {Oppenheim}, {Evans}, \&
  {Martinez-Sykora}}]{Dimant2022.TFBI}
{Dimant}, Y.~S., {Oppenheim}, M.~M., {Evans}, S., \& {Martinez-Sykora}, J.
  2022, arXiv e-prints, arXiv:2211.05264.
\newblock \doarXiv{2211.05264}

\bibitem[{{Dimant} \& {Sudan}(1995)}]{Dimant1995}
{Dimant}, Y.~S., \& {Sudan}, R.~N. 1995, \jgr, 100, 14605,
  \dodoi{10.1029/95JA00794}

\bibitem[{{Fletcher} {et~al.}(2018){Fletcher}, {Dimant}, {Oppenheim}, \&
  {Fontenla}}]{Fletcher2018}
{Fletcher}, A.~C., {Dimant}, Y.~S., {Oppenheim}, M.~M., \& {Fontenla}, J.~M.
  2018, \apj, 857, 129, \dodoi{10.3847/1538-4357/aab71a}

\bibitem[{{Fontenla}(2005)}]{Fontenla2005}
{Fontenla}, J.~M. 2005, \aap, 442, 1099, \dodoi{10.1051/0004-6361:20053669}

\bibitem[{{Fontenla} {et~al.}(2008){Fontenla}, {Peterson}, \&
  {Harder}}]{Fontenla2008}
{Fontenla}, J.~M., {Peterson}, W.~K., \& {Harder}, J. 2008, \aap, 480, 839,
  \dodoi{10.1051/0004-6361:20078517}

\bibitem[{{Golding} {et~al.}(2014){Golding}, {Carlsson}, \&
  {Leenaarts}}]{Golding2014}
{Golding}, T.~P., {Carlsson}, M., \& {Leenaarts}, J. 2014, \apj, 784, 30,
  \dodoi{10.1088/0004-637X/784/1/30}

\bibitem[{{Golding} {et~al.}(2016){Golding}, {Leenaarts}, \&
  {Carlsson}}]{Golding2016}
{Golding}, T.~P., {Leenaarts}, J., \& {Carlsson}, M. 2016, \apj, 817, 125,
  \dodoi{10.3847/0004-637X/817/2/125}

\bibitem[{{Gudiksen} {et~al.}(2011){Gudiksen}, {Carlsson}, {Hansteen}, {Hayek},
  {Leenaarts}, \& {Mart{\'\i}nez-Sykora}}]{Gudiksen2011}
{Gudiksen}, B.~V., {Carlsson}, M., {Hansteen}, V.~H., {et~al.} 2011, \aap, 531,
  A154, \dodoi{10.1051/0004-6361/201116520}

\bibitem[{{Hansteen} {et~al.}(2007){Hansteen}, {de Pontieu}, {Carlsson},
  {McIntosh}, {Watanabe}, {Warren}, {Harra}, {Hara}, {Tarbell}, {Shine},
  {Title}, {Schrijver}, {Tsuneta}, {Katsukawa}, {Ichimoto}, {Suematsu}, \&
  {Shimizu}}]{Hansteen2007}
{Hansteen}, V.~H., {de Pontieu}, B., {Carlsson}, M., {et~al.} 2007, \pasj, 59,
  S699, \dodoi{10.1093/pasj/59.sp3.S699}

\bibitem[{{Hyman}(1979)}]{Hyman1979}
{Hyman}, J.~M. 1979, in Advances in Computer Methods for Partial Differential
  Equations - III, 313--321

\bibitem[{{Leake} {et~al.}(2014){Leake}, {DeVore}, {Thayer}, {Burns},
  {Crowley}, {Gilbert}, {Huba}, {Krall}, {Linton}, {Lukin}, \&
  {Wang}}]{Leake2014}
{Leake}, J.~E., {DeVore}, C.~R., {Thayer}, J.~P., {et~al.} 2014, \ssr, 184,
  107, \dodoi{10.1007/s11214-014-0103-1}

\bibitem[{{Leenaarts} {et~al.}(2007){Leenaarts}, {Carlsson}, {Hansteen}, \&
  {Rutten}}]{Leenaarts2007}
{Leenaarts}, J., {Carlsson}, M., {Hansteen}, V., \& {Rutten}, R.~J. 2007, \aap,
  473, 625, \dodoi{10.1051/0004-6361:20078161}

\bibitem[{{Madsen} {et~al.}(2014){Madsen}, {Dimant}, {Oppenheim}, \&
  {Fontenla}}]{Madsen2014}
{Madsen}, C.~A., {Dimant}, Y.~S., {Oppenheim}, M.~M., \& {Fontenla}, J.~M.
  2014, \apj, 783, 128, \dodoi{10.1088/0004-637X/783/2/128}

\bibitem[{{Mart{\'\i}nez-Sykora} {et~al.}(2012){Mart{\'\i}nez-Sykora}, {De
  Pontieu}, \& {Hansteen}}]{MartinezSykora2012}
{Mart{\'\i}nez-Sykora}, J., {De Pontieu}, B., \& {Hansteen}, V. 2012, \apj,
  753, 161, \dodoi{10.1088/0004-637X/753/2/161}

\bibitem[{{Mart{\'\i}nez-Sykora} {et~al.}(2015){Mart{\'\i}nez-Sykora}, {De
  Pontieu}, {Hansteen}, \& {Carlsson}}]{MartinezSykora2015}
{Mart{\'\i}nez-Sykora}, J., {De Pontieu}, B., {Hansteen}, V., \& {Carlsson}, M.
  2015, Philosophical Transactions of the Royal Society of London Series A,
  373, 20140268, \dodoi{10.1098/rsta.2014.0268}

\bibitem[{{Mart{\'\i}nez-Sykora} {et~al.}(2017){Mart{\'\i}nez-Sykora}, {De
  Pontieu}, {Hansteen}, {Rouppe van der Voort}, {Carlsson}, \&
  {Pereira}}]{MartinezSykora2017}
{Mart{\'\i}nez-Sykora}, J., {De Pontieu}, B., {Hansteen}, V.~H., {et~al.} 2017,
  Science, 356, 1269, \dodoi{10.1126/science.aah5412}

\bibitem[{{Mart{\'\i}nez-Sykora} {et~al.}(2020){Mart{\'\i}nez-Sykora},
  {Szydlarski}, {Hansteen}, \& {De Pontieu}}]{MartinezSykora2020.Ebysus}
{Mart{\'\i}nez-Sykora}, J., {Szydlarski}, M., {Hansteen}, V.~H., \& {De
  Pontieu}, B. 2020, \apj, 900, 101, \dodoi{10.3847/1538-4357/ababa3}

\bibitem[{{N{\'o}brega-Siverio} {et~al.}(2020){N{\'o}brega-Siverio},
  {Mart{\'\i}nez-Sykora}, {Moreno-Insertis}, \&
  {Carlsson}}]{NobregaSiverio2020}
{N{\'o}brega-Siverio}, D., {Mart{\'\i}nez-Sykora}, J., {Moreno-Insertis}, F.,
  \& {Carlsson}, M. 2020, \aap, 638, A79, \dodoi{10.1051/0004-6361/202037809}

\bibitem[{Oppenheim {et~al.}(2020)Oppenheim, Dimant, Longley, \&
  Fletcher}]{Oppenheim2020}
Oppenheim, M., Dimant, Y., Longley, W., \& Fletcher, A.~C. 2020, ApJL, 1, L9

\bibitem[{{Pereira} {et~al.}(2013){Pereira}, {De Pontieu}, \&
  {Carlsson}}]{Pereira2013}
{Pereira}, T. M.~D., {De Pontieu}, B., \& {Carlsson}, M. 2013, \apj, 764, 69,
  \dodoi{10.1088/0004-637X/764/1/69}

\bibitem[{{Przybylski} {et~al.}(2022){Przybylski}, {Cameron}, {Solanki},
  {Rempel}, {Leenaarts}, {Anusha}, {Witzke}, \& {Shapiro}}]{Przybylski2022}
{Przybylski}, D., {Cameron}, R., {Solanki}, S.~K., {et~al.} 2022, \aap, 664,
  A91, \dodoi{10.1051/0004-6361/202141230}

\bibitem[{{Wargnier} {et~al.}(2022){Wargnier}, {Mart{\'\i}nez-Sykora},
  {Hansteen}, \& {De Pontieu}}]{Wargnier2022}
{Wargnier}, Q.~M., {Mart{\'\i}nez-Sykora}, J., {Hansteen}, V.~H., \& {De
  Pontieu}, B. 2022, \apj, 933, 205, \dodoi{10.3847/1538-4357/ac6e62}

\bibitem[{{Wedemeyer} {et~al.}(2004){Wedemeyer}, {Freytag}, {Steffen},
  {Ludwig}, \& {Holweger}}]{Wedemeyer2004}
{Wedemeyer}, S., {Freytag}, B., {Steffen}, M., {Ludwig}, H.~G., \& {Holweger},
  H. 2004, \aap, 414, 1121, \dodoi{10.1051/0004-6361:20031682}

\end{thebibliography}
\bibliographystyle{aasjournal}

\appendix 

\section{Linear Multi-fluid Instability Theory} \label{sec:disprel}
  Starting with fluid equations \eqref{eqs:fluid}, one may derive a theoretical prediction for the properties of linear waves. This is done by linearizing the equations and assuming the original equations hold for the unperturbed values of each quantity. The resulting system of differential equations can be solved by plugging in the ansatz that for some real $\K$ and complex $\W$ all perturbations are proportional to $\exp\brackets{i\parens{\K\cdot\vec{x} - \W t}}$. This yields a linear system of equations in the perturbed quantities. Eliminating the perturbed quantities provides a relationship between $\K$, $\W$, and the unperturbed background.

  In this work we allow for an arbitrary number of ion fluids with arbitrary magnetization, and we include thermal terms. However, we still make some further assumptions to simplify the algebra. In particular, we assume there is only one neutral fluid, $n$, which does not respond to any perturbations, neglect collisional effects between non-neutral fluids (``Coulomb collisions''), and assume all other collision frequencies are constant. We also assume the perturbation is electrostatic, i.e. the magnetic field's response to the perturbation is negligible. Finally, we consider only those solutions where the wavevector $\K$ is perpendicular to the magnetic field $\vec{B}$. After considerable algebra, we find the dispersion relation is:

  \begin{subequations} \label{eqs:disprel}
    \begin{equation} \label{eq:disprel}
      1
      + \sum_{i \in \text{ions}} \parens{\frac{\Ldebye{e}^2}{\Ldebye{i}^2}} \frac{\Fterm_i}{\Fterm_e}
      =
      0   
    \end{equation}
    where the terms are defined as follows:
    \begin{widetext}
      \begin{align}
        \label{eq:Fs}
          \Fterm_\A &\eqdef
          A_\A \parens{
              1
            - \brackets{1 + \frac{2}{3 \mu_\A}} A_\A
            - \frac{B_\A}{\mu_\A}
          }^{-1}
        \\  
        \label{eq:As}
          A_\A &\eqdef
          - i
          \frac{T_\A\oZ}{m_\A}
          \frac{\K \cdot \K}{\nu_{\A n} \W_\A}
          \brackets{
            \frac{\Wterm_\A}{\Wterm_\A^2 + \kappa_\A^2}
          }
        \\  
        \label{eq:Bs}
          B_\A &\eqdef
          \frac{4 m_n}{3 (m_n + m_\A)}
          \frac{\parens{\vec{u}_\A\oZ - \vec{u}_n}}{\W_\A} \cdot 
          \parens{
            \brackets{
              \frac{\Wterm_\A}{\Wterm_\A^2 + \kappa_\A^2}
            } \K
            +
            \brackets{
              \frac{\kappa_\A}{\Wterm_\A^2 + \kappa_\A^2}
            } \frac{\K \times \vec{B}}{|\vec{B}|}
          }
        \\  
        \label{eq:mus}
          \mu_\A &\eqdef
          1 + 2 i \parens{\frac{m_\A}{m_n + m_\A}} \frac{\nu_{\A n}}{\W_\A}
        \\  
          \Wterm_\A &\eqdef
          1 - i \W_\A / \nu_{\A n}
        \\  
        \label{eq:ws in disprel}
          \W_s &\eqdef
          \W - \K \cdot \vec{u}_\A\oZ
      \end{align}
    \end{widetext}

    \noindent and the debye length and magnetization parameter are defined in the usual way:
    \begin{equation} \label{eq:ldebye and kappa}
        \Ldebye{\A}^2 \eqdef
        \frac{
            \epsilon_0 T_\A\oZ
        }{
            q_\A^2 n_\A\oZ
        },
        \quad \quad
        \kappa_\A \eqdef \frac{q_\A |\vec{B}|}{m_\A \nu_{\A n}}
    \end{equation}
  \end{subequations}

  \noindent Above, $n_\A\oZ$, $\vec{u}_\A\oZ$, and $T_\A\oZ$ are the background number density, velocity, and temperature (in energy units) of non-neutral fluid $\A$. By our assumptions, the neutral fluid does not respond to the perturbation, so the neutral velocity $\vec{u}_n$ is constant. Note these expressions adopt the convention $q_e < 0$.

  Through further manipulation, the dispersion relation may be rewritten into a ratio of polynomials in $\W$. We use the author's algebraic manipulation package, \code{SymSolver}\footnote{\url{https://gitlab.com/Sevans7/symsolver}}, to accomplish this task, rather than do it by hand. Considering only two ion species as done in this work, the resulting polynomials are 18th-order in $\W$.
  Such a system is too complicated to solve analytically. However, there are many existing routines for finding the roots of polynomials numerically. In this work, we use the \code{roots} method from the \code{numpy.polynomial} package to find the roots of polynomials numerically for a given set of physical parameters and for each value of $\K$.

  A more detailed derivation and analysis for this dispersion relation and the \TFBI\ theory can be found in \citet{Dimant2022.TFBI}. Note there are a couple differences between the dispersion relation here and in that work. Here, we include the case where neutral velocity is nonzero. Also, since \citet{Dimant2022.TFBI} uses the poisson equation instead of assuming quasineutrality, there is an additional term which appears on the right hand side of the generic dispersion relation in that work. For the parameters of the simulation in our work, we confirmed numerically that this additional term has negligible effect on growth rate predictions.

\section{Numerical Scheme --- Artificial Diffusion} \label{sec:hyperdiffusion}
  The artificial hyperdiffusion terms in Ebysus primarily diffuse sharp fluctuations at small scales (5 grid cells or less). These terms are similar to those in Bifrost \citep{Gudiksen2011}, but have been adapted to the multi-fluid simulations discussed in this work. In particular, these terms are added to the continuity, momentum (each component treated separately), and energy equations (\ref{eq:continuity}, \ref{eq:momentum}, \ref{eq:heating}) for every fluid. Including only the hyperdiffusion terms used in this work, each of these equations $\partial f_\A/\partial t = \text{\textit{(original RHS)}}$ becomes:
  \begin{widetext}
    \begin{subequations}
      \begin{equation}\label{eq:hyperdiff}
        \pdtime{f_\A} =
          \text{\textit{(original RHS)}}
          + \nu_1 C_\A^{\text{(fast)}}
          \sum_{x'_i \in \{x', y'\}}
          \pdfrac{}{x'_i}
          \brackets{
            \pdfrac{f_\A}{x'_i}
            Q_{x'_i}\parens{\pdfrac{f_\A}{x'_i}}
          }
      \end{equation}
      where
      \begin{equation}
        Q_{x'_i}(g) = 
          \left.
          \parens{\pdfrac{}{x'_i}\pdfrac{g}{x'_i}}
          \middle/
          \parens{|g| + \frac{1}{q}\pdfrac{}{x'_i}\pdfrac{g}{x'_i}}
          \right.
      \end{equation}
    \end{subequations}
  \end{widetext}
  Here, $\nu_1=0.01$ and $q=1.0$ are constants, $C_\A^{\text{(fast)}} = \sqrt{\parens{\gamma T_\A / m_\A}^2 + B^2 / \parens{\mu_0 m_\A n_\A}}$ is the speed of the fast magnetosonic wave for fluid $\A$, $\gamma = 5/3$, and $x'$ and $y'$ are the spatial coordinates $x$ and $y$ normalized such that grid cells each have length 1.

\section{Simulation Electric Field} \label{sec:E vs t}
\begin{figure}  
      \includegraphics[width=\customhalfwidth]{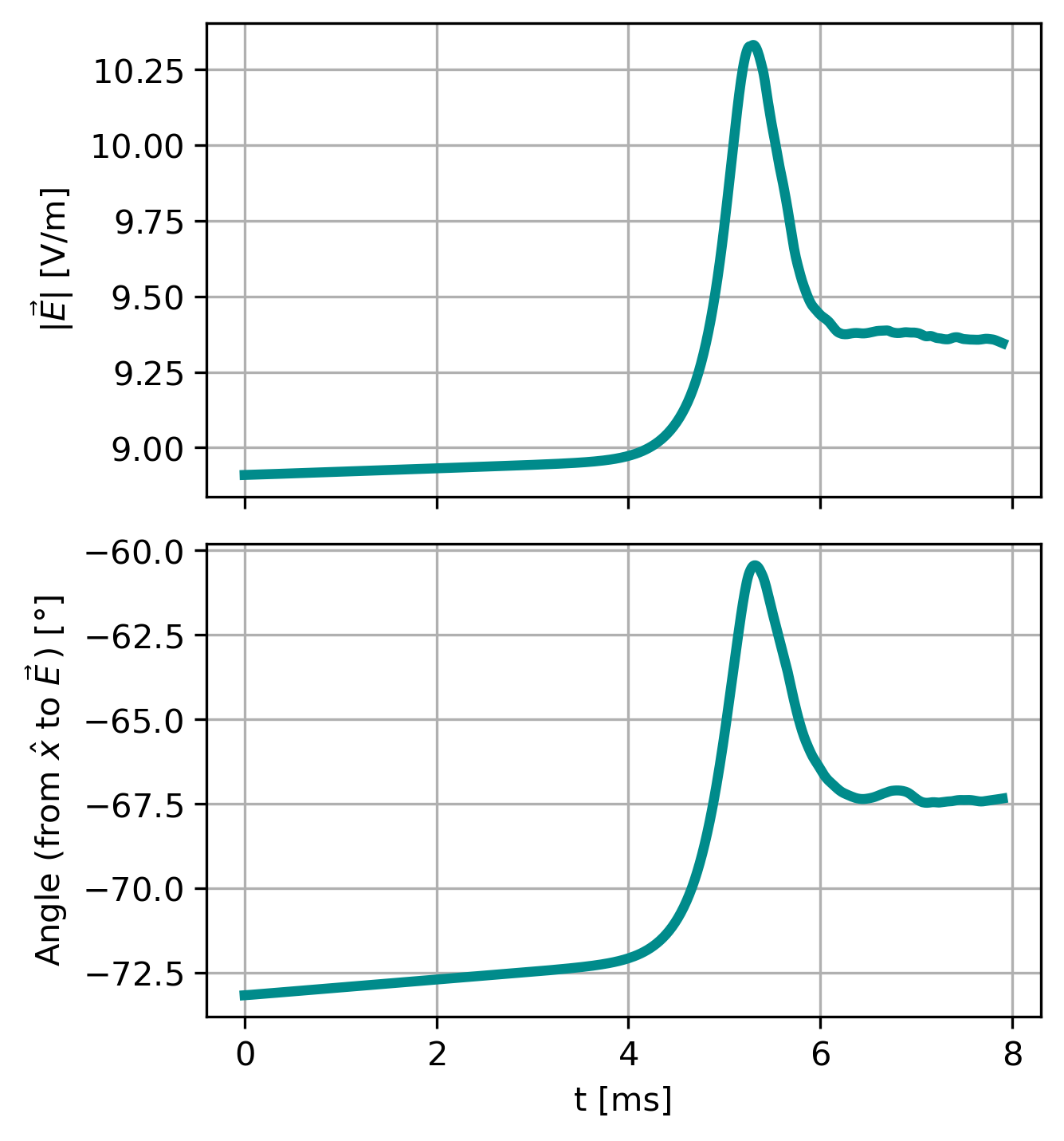}
      \caption{
        \label{fig:E vs t}
        Mean electric field throughout the simulation of the \TFBIabbrv. The top plot indicates the magnitude, while the bottom plot indicates the angle with respect to the positive x-axis.
      }
    \end{figure}

  The electric field varies throughout the Ebysus simulations discussed in this work. Figure~\ref{fig:E vs t} plots the mean electric field for the main simulation of the \TFBI\ presented in this work, see for example Section~\ref{sec:simulation results} and Figure~\ref{fig:simulation}. This figure shows the magnitude and direction of the mean electric field throughout the simulation, calculated by solving the electron momentum equation for $\vec{E}$ assuming no electron inertia. The magnitude increases roughly linearly from $8.91$ to $8.97$~V/m during the first $4.0$~ms of the simulation, increases to its peak of $10.33$~V/m at $t=5.30$~ms, decreases, then fluctuates before reaching its final value of $9.35$~V/m at the end of the simulation. The angle increases roughly linearly from $-73.2\degr$ to $-72.1\degr$ during the first $4.0$~ms, increases to its peak of $-60.4\degr$ at $t=5.32$~ms, decreases, then fluctuates before reaching its final value of $-67.3\degr$.

\end{document}